\documentclass[aps,prb,twocolumn,superscriptaddress,noeprint,longbibliography, nofootinbib,floatfix]{revtex4-2}

\usepackage{graphicx}
\usepackage{bm}
\usepackage{amsmath, amssymb}
\usepackage{booktabs}
\usepackage{mathrsfs}
\usepackage{makecell}
\usepackage[normalem]{ulem}
\usepackage{pifont}
\usepackage{placeins}
\usepackage[mathscr]{euscript}
\usepackage{physics}
\usepackage{xcolor}
\usepackage{siunitx}
\usepackage{subcaption}
\usepackage{appendix}
\usepackage{mathtools}
\usepackage[%
  pdfstartview=FitH,%
  breaklinks=true,%
  bookmarks=true,%
  colorlinks=true,%
  anchorcolor=black,%
  citecolor=blue,
  filecolor=black,%
  menucolor=black,%
  urlcolor=blue,%
  linkcolor=blue,%
 ]{hyperref}
 \usepackage[format=plain,
      justification=RaggedRight,
      singlelinecheck=false]
     {caption}
\usepackage{subcaption}
\usepackage[normalem]{ulem}
\newcommand{\stkout}[1]{\ifmmode\text{\sout{\ensuremath{#1}}}\else\sout{#1}\fi}
\graphicspath{ {./Figures/} }

\newcommand{\cf}{\textit{cf.} }

\newcommand{\eg}{\textit{e.g.}}

\newcommand{\SL}{\mathcal{S}_{[111]}}
\newcommand{\SLa}{\mathcal{S}_k}

\def\equationautorefname~#1\null{Eq. (#1)\null}

\newcommand{\appref}[1]{\hyperref[#1]{App.~\ref*{#1}}}
\usepackage[all]{hypcap} %
\usepackage{physics}

\newcommand{\taulock}{\tau_{\mathrm{lock}}}
\usepackage[backgroundcolor=green!40,textsize=tiny]{todonotes}

\begin{document}

\title{[111]-strained spin ice: Localization of thermodynamically deconfined monopoles}


\author{Zhongling Lu}
\email{zhongling\_lu@sjtu.edu.cn}
\address{Department of Physics, Boston University, Boston, Massachusetts 02215, USA}
\address{Zhiyuan College, Shanghai Jiao Tong University, Shanghai 200240, China}

\author{Robin Sch{\"a}fer}
\email{rschaefe@bu.edu}
\address{Department of Physics, Boston University, Boston, Massachusetts 02215, USA}

\author{Jonathan N. Hall\'en}
\email{jonathan\_nilssonhallen@g.harvard.edu}
\address{Department of Physics, Boston University, Boston, Massachusetts 02215, USA}
\address{Department of Physics, Harvard University, Cambridge, Massachusetts 02138, USA}

\author{Chris R. Laumann}
\email{claumann@bu.edu}
\address{Department of Physics, Boston University, Boston, Massachusetts 02215, USA}
\address{Department of Physics, Harvard University, Cambridge, Massachusetts 02138, USA}
\address{Max-Planck-Institut f\"{u}r Physik komplexer Systeme, 01187 Dresden, Germany}

\date{\today}

\begin{abstract}
We study classical spin ice under uniaxial strain along the $[111]$ crystallographic axis. 
Remarkably, such strain preserves the extensive ice degeneracy and the corresponding classical Coulomb phase.
The emergent monopole excitations remain thermodynamically deconfined exactly as in the isotropic case.
However, their motion under local heat bath dynamics depends qualitatively on the sign of the strain.
In the low-temperature limit 
for negative strain, the monopoles diffuse, 
while for positive strain, they localize. 
Introducing additional ring exchange dynamics into the ice background transforms the localized monopoles into subdimensional excitations whose motion is restricted to diffusion in the (111)-plane.
The phenomena we identify are experimentally accessible in rare-earth pyrochlores under uniaxial stress as well as in tripod kagome materials.
The diffusive versus localized nature of the monopoles manifests in characteristic magnetic noise spectra, which we compute.
\end{abstract}

\maketitle

\section{Introduction \label{sec:intro}}
\begin{figure}[t]
    \begin{subfigure}[h]{0.45\columnwidth}
    \centering
    \includegraphics[width=\textwidth]{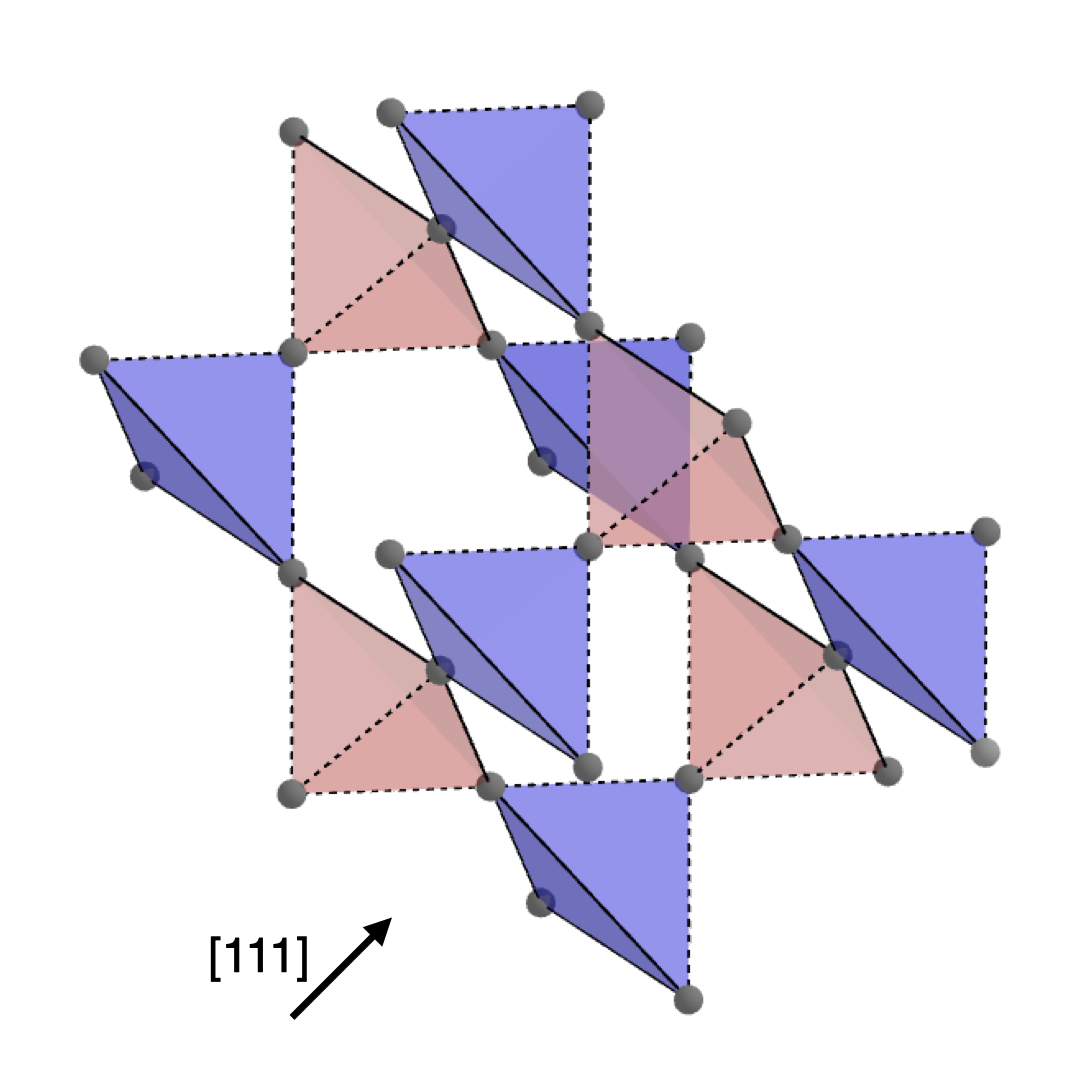}
    \caption{Uniaxial stress.}
    \label{fig:pyrochlore_pressure}
    \end{subfigure}
    \begin{subfigure}[h]{0.45\columnwidth}
    \centering
    \includegraphics[width=\textwidth]{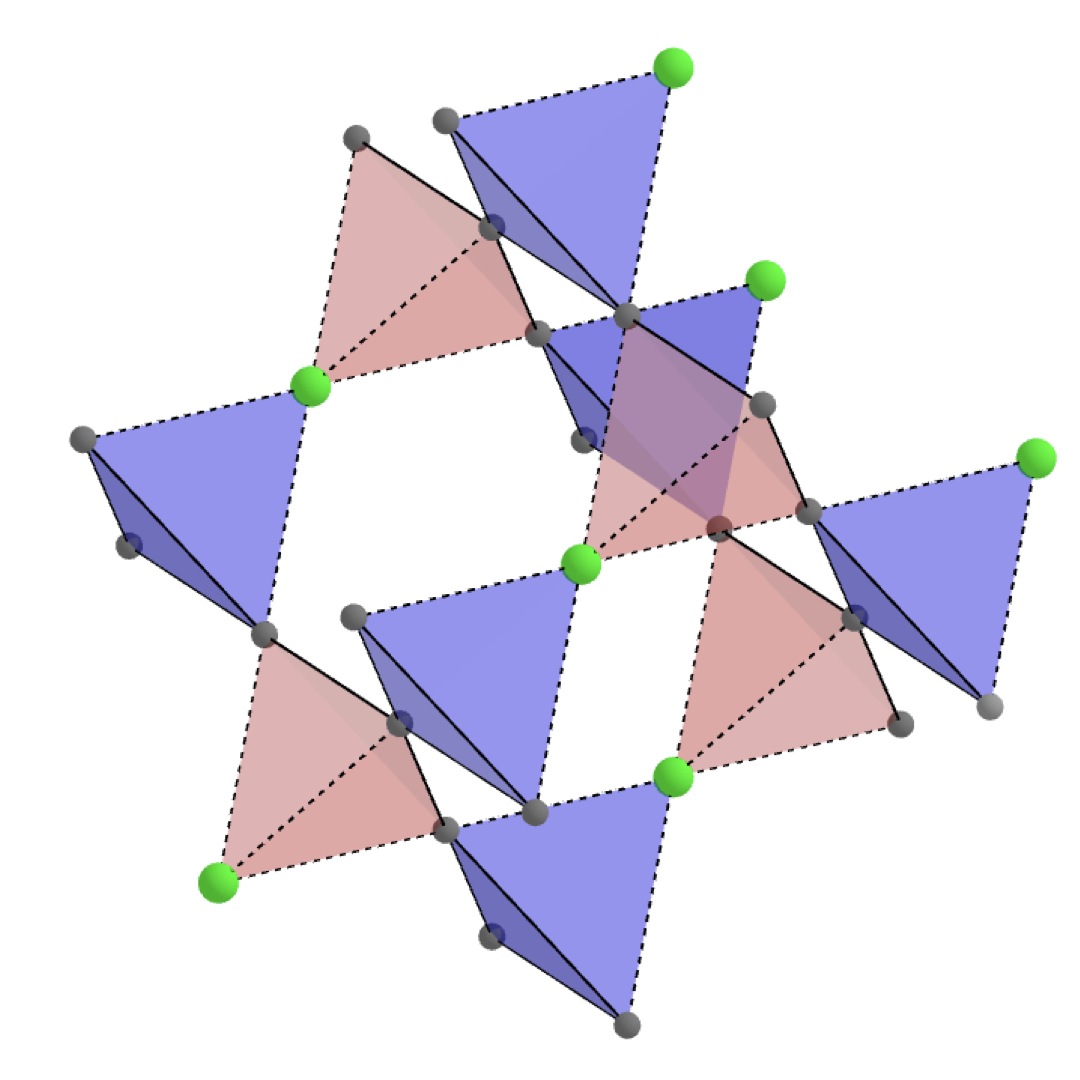}
    \caption{Sublattice replacement.}
    \label{fig:pyrochlore_replaced}
    \end{subfigure}
    \caption{Two illustrations of the [111]-deformed pyrochlore with symmetry-reduced space group $R\Bar{3}m$ describing the nearest-neighbor spin ice model in  \autoref{eq:Hamiltonian_nn}. Distorted bonds along the $[111]$ direction are highlighted with dashed lines. 
    (a) Strain induced by uniaxial stress where bonds along the $[111]$ direction are squeezed, increasing the interaction strength.  
    (b) Substitution of the atoms located on the $[111]$-sublattice (green)  results in a symmetry-equivalent modulation of the nearest-neighbor interaction. 
    Tripod kagome systems~\cite{dun_magnetic_2016,paddison_emergent_2016} realize the extreme case where the green atoms are nonmagnetic and the kagome planes decouple.
    }
    \label{fig:pyrochlores}
\end{figure}

The nearest-neighbor Ising model on the pyrochlore lattice, called \textit{spin ice}, is the quintessential {three-dimensional} frustrated magnetic system~\cite{anderson1956ordering, bramwell2020history, udagawa2021spin}. 
Its unusual properties include extensive zero-point entropy~\cite{anderson1956ordering, singh_correction_2012} and fractionalized excitations in the form of \textit{thermodynamically deconfined} magnetic monopoles~\cite{Castelnovo2008, castelnovo2012spin}. 
This work focuses on spin ice under uniaxial $[111]$-strain; see \autoref{fig:pyrochlores}.
Such strain preserves the extensive entropy corresponding to the thermodynamic Coulomb phase. 
Nonetheless, the strain leads to dramatic changes in the \emph{dynamics} of the monopoles.

In the minimal, ``standard model'' of spin ice dynamics, single spins flip stochastically according to a Markovian heat-bath~\cite{Jaubert2011}{, which} leads to monopole diffusion in the isotropic case~\cite{hallen2022}. 
Remarkably, the introduction of $[111]$-strain leads to qualitatively distinct monopole dynamics depending on the sign of the strain. 
If the strain increases the interactions between spins separated in the $[111]$-direction, the monopoles completely localize.
In the opposite scenario, monopoles remain diffusive with a weak anisotropic correction to the diffusion tensor. 
Both the dynamical localization and the anisotropic diffusion manifest in macroscopic dynamical properties such as the magnetic noise spectrum~\cite{dusad2019magnetic, samarakoon2022anomalous}, which has been of much recent experimental interest. 
{Moderate amounts of strain preserve the ice degeneracy of all two-in-two-out states, and the monopoles remain thermodynamically deconfined excitations but become \textit{dynamically localized}.
By this, we mean that a monopole excitation can be moved any distance at a finite, or even zero, energy cost; however, in the low-temperature limit, the timescale on which a monopole moves even a short distance diverges.}

If one additionally includes {fluctuations in} the ice background on top of the standard model dynamics, then the localized monopoles become diffusive in the $(111)$-plane, leading to \textit{subdimensional} diffusion.
In this sense, they are classical ``subdimensional excitations''~\cite{nandkishore2019fractons,pretko2020fracton}.

%
%
\begin{figure*}[ht!]
    \centering
    \includegraphics[width=1.\textwidth]{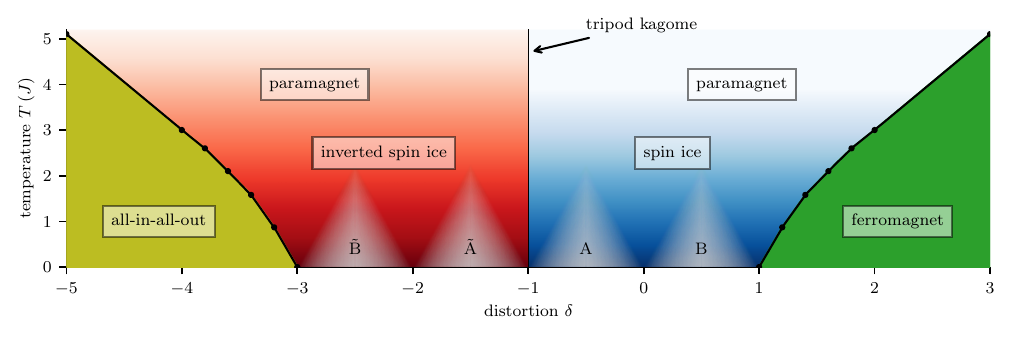}
	\caption{The phase diagram of strained spin ice, described by the Hamiltonian $H_{nn}$ (\autoref{eq:Hamiltonian_nn}), as a function of distortion, $\delta$, and temperature, $T\,(J)$. The model realizes two ordered and two disordered low-temperature phases that are connected by the duality axis at $\delta = -1$. For $\delta < -3$, the model is long-range ordered following an \textit{all-in-all-out} pattern (yellow). Under duality, \autoref{eq:transformation}, the ordered state maps to a ferromagnetic state at $\delta > 2$ (green). Similarly, the duality maps the famous \textit{spin ice} phase (blue) for $-1 < \delta < 1$ to an equivalent disordered Coulomb phase (red) we name \textit{inverted spin ice} for $-3 < \delta < -1$. 
    The model changes character at $\delta = -1$, the duality axis, which marks a transition from ferro- to antiferromagnetic couplings along $[111]$. At the duality point, $[111]$ spins decouple, and the system is composed of alternating antiferromagnetic kagome planes sandwiched between planes of free spins.
    The white-shaded regions indicate the temperature regimes where the dynamics are strongly influenced by strain; in the A regimes, monopoles are (anisotropically) diffusive, whereas they are dynamically localized in the B regimes.
    Again, the duality provides the analogous dynamical description in the inverted spin ice phase. 
    }
    \label{fig:PD_strain}
\end{figure*}
%
%

Our model is experimentally motivated, and moderate values of strain can be induced by stress along the $[111]$ crystallographic axis, see \autoref{fig:pyrochlore_pressure}. 
Experiments on the classical spin ice compounds Dy$_2$Ti$_2$O$_7$ and Ho$_2$Ti$_2$O$_7$ have been carried out with uniaxial stress applied along the $[100]$, $[110]$, and $[111]$ axes~\cite{mito2007uniaxial, edberg2019dipolar, edberg2020effects, pili2022topological}. 
These experiments probed static quantities, 
which are insensitive to the monopole localization we predict.

Atomic substitution is an alternate route to exploring spin ice with the same reduced symmetry as the $[111]$-strained pyrochlore.
By systematically substituting atoms on a particular sublattice, one obtains larger effective strain parameters, see \autoref{fig:pyrochlore_replaced}.
This could even flip the sign of the interactions between different species relative to the usual spin ice.
A special case of such materials is provided by kagome Ising magnets such as Dy$_3$Mg$_2$Sb$_2$O$_{14}$~\cite{dun_magnetic_2016,paddison_emergent_2016}. 
These realize the extreme case where one sublattice spin is nonmagnetic, leaving effectively decoupled kagome layers.
 
Motivated by this possibility, we compute the full phase diagram of the symmetry-reduced model, shown in \autoref{fig:PD_strain}. 
Using exact zero-temperature analysis combined with a lattice-scale duality transformation, we identify two distinct Coulomb phases and two distinct symmetry-breaking ordered phases. 
We further confirm the finite-temperature extent of these phases using Monte Carlo techniques.

Thermodynamic properties of spin ice under uniaxial strain have been investigated theoretically in the past~\cite{jaubert2010spin, xie2015magnetic}. However, the main focus of previous works was the effect of stress applied along the $[100]$ direction, where any amount of strain breaks the extensive degeneracy of the ice states, resulting in qualitatively different behavior compared to the $[111]$ case we consider.

The remainder of the paper is structured as follows:
\autoref{sec:GS and excitations} introduces the model and the exact ground states in the four thermodynamic phases, as well as the relevant excitations. \autoref{sec:thermo} then explores the finite-temperature thermodynamics of our model using Monte Carlo simulations, and \autoref{sec:dynamics} is devoted to the low-temperature equilibrium dynamics of the monopoles.
Possible experimental realizations, including estimates of realistic values of the distortion, are addressed in \autoref{sec:experiments}.

\section{Ground states and excitations \label{sec:GS and excitations}}
\begin{figure*}[t]
    \centering
    \includegraphics[width=0.57\textwidth]{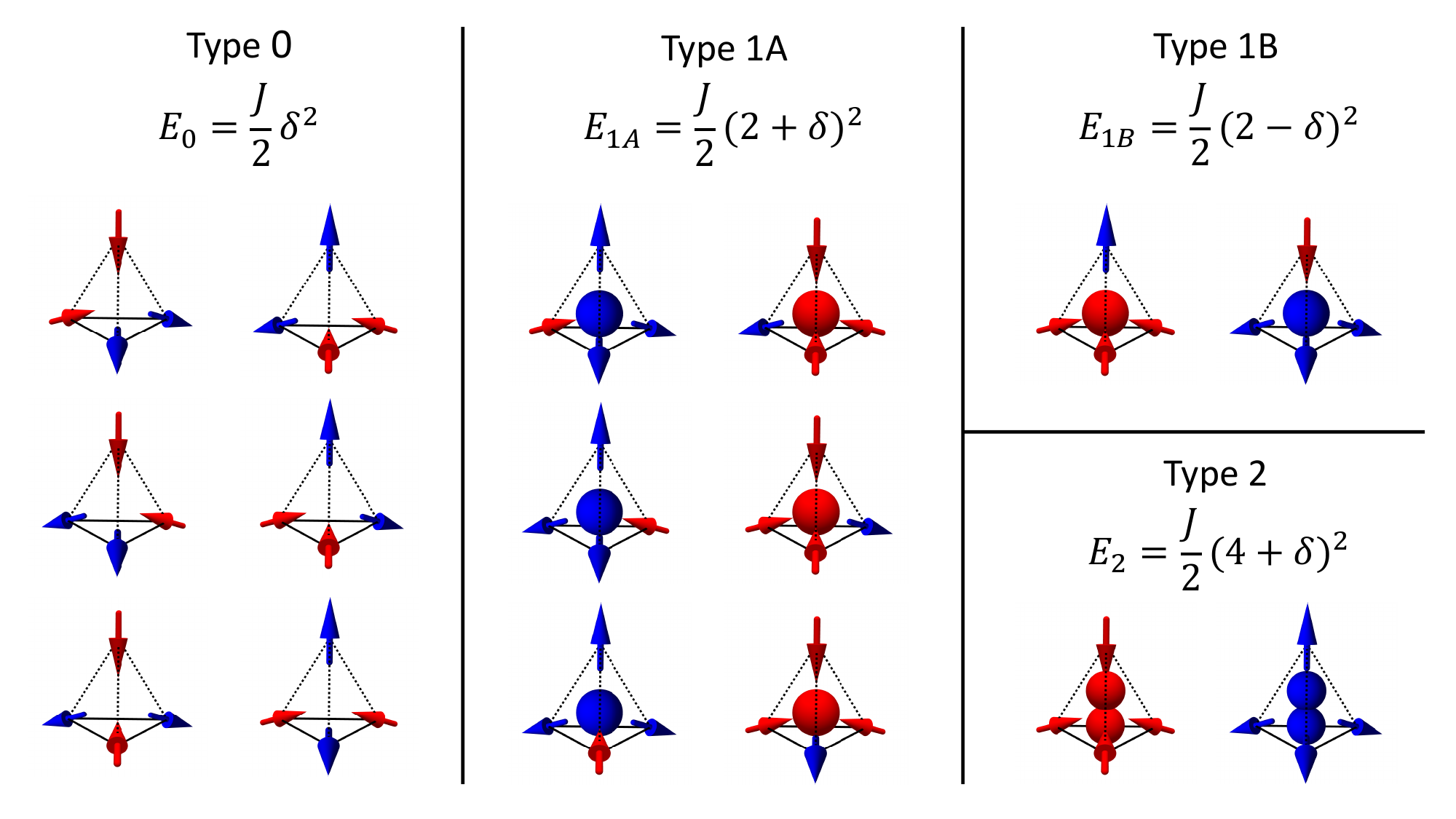}
    \hfill
    \includegraphics[width=0.42\textwidth]{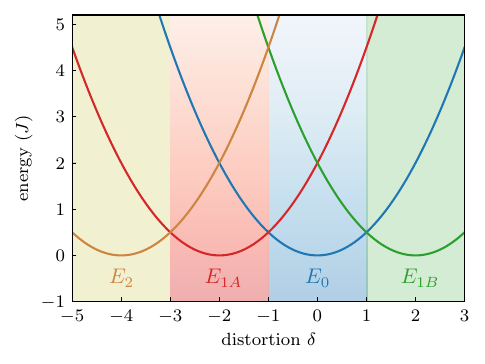}
    \caption{\textit{Left}: All 16 spin configurations on a single tetrahedron organized into four degenerate groups according to $H_S$, \autoref{eq:Hamiltonian}. 
    The energies of the groups depend on the strain parameter, $\delta$, and each phase in \autoref{fig:PD_strain} corresponds to the regime where energy is minimized by tiling the system with configurations from a particular group.
    Type 0 and type 1A configurations, respectively, correspond to the spin ice and inverted spin ice phases, with an extensive number of possible ground states. In contrast, the ferromagnetic (type 1B) and the all-in-all-out (type 2) phases are long-range ordered with only two ground state configurations each.
    The duality transformations \autoref{eq:transformation} is a unique mapping between type 0 and type 1A as well as type 1B and type 2 configurations.
    Monopoles and anti-monopoles are highlighted in blue and red, respectively. The sublattice $\SL$ spin is located on the top of the tetrahedra and couples to the three $\SLa$ spins with interaction strength $J(1+\delta)$, illustrated by dashed bonds. The three $\SLa$ spins interact with strength $J$, visualized by solid lines.
    \textit{Right}: Energy versus $\delta$ for the four configuration groups. The colors of the lines and background correspond to the phases in \autoref{fig:PD_strain}. It further highlights the most relevant excitation type at each value of $\delta$. For example, within the spin ice phase, the type 1A monopoles (red) dominate for $-1 < \delta < 0$, and type 1B monopoles (green) dominate for $0<\delta < 1$. 
    }
    \label{fig:configs}
\end{figure*}

Our starting point is the only symmetry-allowed classical nearest-neighbor Hamiltonian describing spin ice under strain:
\begin{equation}\label{eq:Hamiltonian_vector}
    H_{\rm nn} = -3J (1 + \delta) \sum_{\mathclap{\substack{\langle i, j \rangle \\ i \in \SL,\ j \notin \SL}}} \vec{s}_i \cdot \vec{s}_j - J \sum_{\mathclap{\substack{\langle i, j \rangle \\ i,j \notin \SL}}} \vec{s}_i \cdot \vec{s}_j \, .
\end{equation}
where $\delta$ is a dimensionless scalar parameter that modulates the interaction strength and can take both positive and negative values. The first sum runs over pairs with one spin on sublattice $\SL$ -- the sublattice on which spins align with the $[111]$ direction -- and the second sum runs over pairs with no spin on sublattice $\SL$. We denote the other three sublattices collectively as $\SLa$ as these spins form kagome planes. 
{The variables $\vec{s}$ are classical vector spins forced to point along the local $\langle 111 \rangle$ axes as illustrated in ~\autoref{fig:configs}. These can be expressed in terms of Ising variables $\sigma_i=\pm 1$ as $\vec{s}_i = \sigma_i \vec{e}_i$, where $\vec{e}_i$ is the corresponding unit vector, allowing us to rewrite the Hamiltonian as (see \appref{app:spin_notation})
\begin{equation}
    H_{\rm nn} = J (1 + \delta) \sum_{\mathclap{\substack{\langle i, j \rangle \\ i \in \SL,\ j \notin \SL}}} \sigma_i \sigma_j + J \sum_{\mathclap{\substack{\langle i, j \rangle \\ i,j \notin \SL}}} \sigma_i \sigma_j \, .
     \label{eq:Hamiltonian_nn}
\end{equation}
We take $J>0$ as the (ferromagnetic) interaction strength, which corresponds to an antiferromagnetic interaction in the language of the Ising variables.}
A nonzero $\delta$ reduces the space group of the lattice from the usual $Fd\Bar{3}m$ of the pyrochlore lattice to $R\Bar{3}m$ \cite{aroyo2006bilbao, aroyo2006bilbaoII, aroyo2011crystallography, paddison_emergent_2016}.
As in conventional spin ice, we can re-express $H_{\rm nn}$ up to a constant term as a sum over all tetrahedra,
\begin{equation}
    H_S = \frac{J}{2}\sum_t \left((1+\delta)\sigma_{t,1}+\sigma_{t,2}+\sigma_{t,3}+\sigma_{t,4}\right)^2\, ,
    \label{eq:Hamiltonian}
\end{equation}
where $\sigma_{t, 1}$ are members of $\SL$.

%
%

Much of the model can be understood by examining the $2^4 = 16$ spin configurations on a single tetrahedron, see \autoref{fig:configs}. 
Six of these have two spins pointing in and two spins pointing out of the tetrahedron;
we refer to these states as \textit{Type 0} configurations. 
Any configuration where every tetrahedron follows this ``two-in-two-out'' rule, also known as the ``ice rule''~\cite{bernal1933}, is a ground state of isotropic nearest-neighbor spin ice.
As pointed out by Anderson~\cite{anderson1956ordering}, there is an extensive entropy of such ground states, originally estimated by Pauling~\cite{pauling1935} in the context of water ice~\cite{giauque_entropy_1936}. 
Flipping a single spin out of a two-in-two-out ground state creates a pair of excitations that take the form of three-in-one-out and one-in-three-out tetrahedra. 
These are the famous emergent magnetic monopoles~\cite{Castelnovo2008}, which we call \textit{Type 1} configurations. 
Once created, monopoles can move independently through further spin flips and only annihilate if they meet a monopole of the opposite sign. 
Eight of the sixteen spin configurations on a tetrahedron correspond to a single monopole. 
The remaining two ``double monopole'' configurations have all four spins pointing either in or out, and we call these \textit{Type 2} configurations. 

While type 0 and type 2 configurations remain degenerate, the $[111]$-strain lifts the degeneracy of the eight type 1 configurations. 
The single monopoles are type 1A with energy $J(2 + \delta)^2 /2$ if the minority spin is not on $\SL$, and type 1B with energy $J(2 - \delta)^2 /2$ if the minority spin is on $\SL$.
The minority spin is the \textit{out} spin in a {three-in-one-out} tetrahedron and the \textit{in} spin in a {three-out-one-in} tetrahedron.
The two type 1B configurations with minority spin on sublattice $\SL$ are energetically ``cheaper'' for $\delta > 0$, and monopoles with the minority spin on any of the other sublattices are favored for $\delta < 0$.
Monopoles can change type when they move, but this carries an associated energy cost. Dynamical localization of monopoles occurs when the temperature is low compared to the cost of converting a type 1B to a type 1A monopole, as further explained in~\autoref{sec:dynamics}. 

\paragraph*{Duality.} 
Before discussing the different ground states, we focus on the point $\delta=-1$, which reveals a duality of the Hamiltonian $H_S$. 
This can be seen by defining $\eta = 1 + \delta$ and rewriting the Hamiltonian as
\begin{equation}
    H_S(\eta) = \frac{J}{2} \sum_t \left(\eta \sigma_{t,1}+\sigma_{t,2}+\sigma_{t,3}+\sigma_{t,4}\right)^2\label{eq:Hamiltonian_eta}
    \, .
\end{equation}
 At $\eta = 0$ ($\delta = -1$), spins on sublattice $\SL$ are decoupled from their neighbors such that the system is decomposed into a set of kagome planes and a set of completely free spins. 
This phenomenon can be observed in so-called ``tripod kagome'' materials such as Dy$_3$Mg$_2$Sb$_2$O$_{14}$~\cite{dun_magnetic_2016,paddison_emergent_2016}.
We can define a duality transformation, $H_S(\eta)\rightarrow H_S(-\eta)$, by
\begin{equation}
\begin{split}
    \sigma_i &\rightarrow -\sigma_i \text{ for } i\in\SL\\
    \sigma_j &\rightarrow \sigma_j \text{ for } j \in \SLa 
    \, ,
    \label{eq:transformation}
\end{split}
\end{equation}
which is an inversion of all spins on sublattice $\SL$. Utilizing this duality reduces our focus to only half of the phase diagram at $\delta \geq -1$. 
The model is thus symmetric around $\eta=0$, as reflected in the phase diagram,~\autoref{fig:PD_strain}.
We shall now explain the four phases from the right (positive $\delta$) to the left (negative $\delta$).

\paragraph*{Ferromagnetic regime.}
For $\delta > 1$, the energy of the type 1B monopoles $E_{1B} = J(2-\delta)^2/2$ is lower than the energy of any other group.
Ground states are formed by a complete tiling of type 1B monopoles on every tetrahedron with all minority spins on sublattice $\SL$.
There are only two configurations that minimize the energy on all tetrahedra simultaneously, corresponding to polarizing all spins along or against the $[111]$ direction.
The system undergoes a phase transition into this spontaneous symmetry-broken ferromagnetic state, where the transition temperature $T_c$ grows with increasing $\delta$. Phrased differently, this is a stress-induced crystallization of a cooperative paramagnet, akin to phenomena observed in Tb$_2$Ti$_2$O$_7$ \cite{mirebeau2002pressure}.
The ferromagnetic phase persists up to arbitrarily large $\delta$.
The lowest-energy excitations in this regime are two-in-two-out (type 0) states. A single spin flip out of the ground state can create a pair of such defects, but they are not mobile -- there are no deconfined excitations in the ferromagnetic phase.

\paragraph*{Spin ice regime.}
For $\vert\delta\vert<1$, the lowest energy is $E_0=J\delta^2/2$ and the ground states are given by the conventional spin ice configurations, realizing a finite residual entropy at $T=0$~\cite{pauling1935,anderson1956ordering,ramirez_zero-point_1999}. Either the type 1A or type 1B monopoles are the lowest-energy excitations, depending on whether $\delta$ is smaller or greater than zero. 
At $\delta=0$, conventional spin ice, the two types become degenerate, and monopoles diffuse isotropically.

\paragraph*{High symmetry point.}
At the point $\delta=-1$, spins on sublattice $\SL$ are completely decoupled from all other spins.
At this special point, the $\SL$ spins are free to take any value they want, and the remaining spins form decoupled antiferromagnetic kagome planes. In terms of the single tetrahedron, $E_0=E_{1A} = J/2$, making this the most degenerate point, with 12 of the 16 configurations sharing the same energy.

\paragraph*{Inverted spin ice regime.}
For $-3 < \delta < -1$, the energy is minimized by placing a single type 1A monopole on each tetrahedron. Type 1A configurations have a majority spin on sublattice $\SL$, resulting in much more freedom in how to form ground states compared to the type 1B tiled ferromagnetic phase.
In fact, there is a direct one-to-one mapping between the monopole-favoring ground states for $-3 < \delta < -1$ and the two-in-two-out ground states for $-1 < \delta < 1$ due to the duality around $\delta=-1$.
Any ice configuration can be mapped uniquely by the duality transformation, \autoref{eq:transformation}, to a ground state for $-3 < \delta < -1$, and we dub this phase \textit{inverted spin ice}.

\paragraph*{All-in-all-out regime.}
For $\delta < -3$, the double monopole (type 2) energy is the lowest of the four groups and the system orders into an \textit{all-in-all-out} state.
There are two such states with a broken tetrahedral sublattice symmetry, where all spins pointing either from \textit{down} to \textit{up} tetrahedra or from \textit{up} to \textit{down} tetrahedra. In the language of Ising variables, this refers to all $\sigma_i$ being $+1$ or $-1$.
This ordered phase persists for arbitrarily negative $\delta$, and again leads to a symmetry-breaking phase transition at a critical temperature $T_c$ that grows with decreasing $\delta<-3$. 
As before, the duality transformation maps the \textit{all-in-all-out} states to the two ferromagnetically ordered ground states at $\delta > 1$.

%
%
\begin{figure*}[t]
    \centering
    \includegraphics[width=\textwidth]{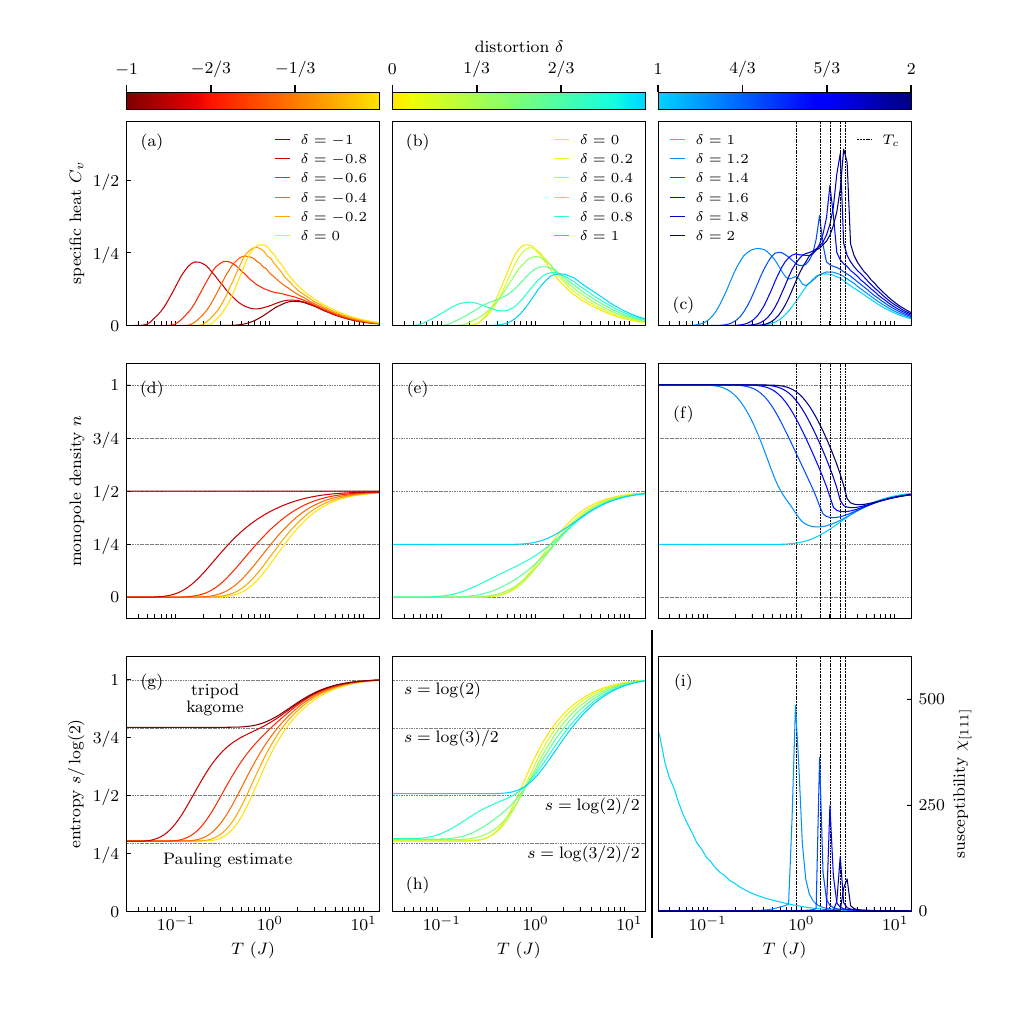}
    \caption{The combined thermodynamic data for $H_S$ in \autoref{eq:Hamiltonian} with $\delta \geq -1$. The left and central columns cover $\delta$ values that correspond to the spin ice regime, whereas the right column refers to the ferromagnetic region at $\delta > 1$. 
    \textit{Top row}: Panels (a,b,c) display the specific heat with the characteristic Schottky anomalies associated with different excitations shown in \autoref{fig:configs}. Within the spin ice phase, panels (a,b), the two distinct peaks are linked to nondegenerate monopole excitations, which collapse into a single smooth peak for degeneracies occurring at $\delta = 0,\pm 1$.
    In the ferromagnetic regime, the phase transition manifests as a sharp peak marked by dotted vertical lines. 
    \textit{Middle row}: Single monopole densities, including both type 1A and 1B, but not counting double monopoles. As the temperature is reduced, the monopole density goes to zero in the spin ice regime, panels (d,e), and to one in the ferromagnetic regime, panel (f).  For the special points, $\delta=\pm 1$, single monopoles and ice states share the same energy, and the monopole density goes to $1/2$ and $1/4$, respectively. 
    \textit{Bottom row}: The entropy per spin is shown for the spin ice regime (g,h), matching Pauling estimates as described. Panel (i) of this row displays the static susceptibility measured in the $[111]$ direction for $\delta> 1$, showing sharp peaks due to the phase transition.}
    \label{fig:thermo}
\end{figure*}
%
%

\section{Thermodynamics \label{sec:thermo}}
We now consider the thermodynamics of our model using classical Monte Carlo simulations.
Similar to the case of isotropic spin ice, we find a cross-over from the disordered Coulomb phases to the paramagnet at finite temperature. In contrast, for the symmetry-breaking phases, we identify a critical temperature marking the transition from the paramagnet into the ordered regime.
We only conduct simulations at $\delta \geq -1$ ($\eta \geq 0$) due to the duality in \autoref{eq:transformation}.
Our results are summarized in \autoref{fig:thermo}, which contains the specific heat (a-c), monopole density (d-f), and entropy (g,h) for the different regimes discussed previously.
Panel (i) shows the magnetic susceptibility measured in the $[111]$ direction.
The first column (a,d,g) covers the parameter regime $-1\leq \delta \leq 0$ -- the region of the spin ice phase where type 1A monopoles are the lowest-energy excitations. 
Similarly, the second column (b,e,h) covers $0\leq \delta \leq 1$ -- the spin ice phase where type 1B monopoles are lowest in energy. 
Lastly, the third column (c,f,i) focuses on the ferromagnetic phase, with clear signs of a phase transition occurring at $T_c$.

\paragraph*{Specific heat.}
The specific heat, $C_V$, contains direct information on the distribution of tetrahedra between the four groups in \autoref{fig:configs}.
Let us start by examining the spin ice phase shown in panels (a,b) of \autoref{fig:thermo}. At infinite temperature, all configurations are equally probable. Upon cooling, higher energy configurations are frozen out. At the level of individual tetrahedra, this means that double monopoles are avoided first, followed by the two types of monopoles. Freezing out these configurations induces a Schottky anomaly, which is visible in the specific heat. When type 1A and 1B monopoles are well separated in energy, each group gives rise to an individual Schottky peak, and as $\vert \delta\vert$ increases, we observe the single specific heat peak of $\delta=0$ splitting into two distinct peaks. 
Note that the entropy released by freezing out these excitations is constant and the heights of the peaks are related to this entropy.
As there are more type 1A than type 1B monopoles, the freezing out of type 1A monopoles leads to a larger peak in the specific heat. As $E_{1A}<E_{1B}$ for $\delta < 0$, the low-temperature peak in the specific heat is higher in this regime, whereas the opposite is true for $\delta > 0$.
At $\delta=-1$ ($\delta=1$), type 1A (1B) monopoles are degenerate with the type 0 configurations, and one of the two monopole types is not frozen out, leaving a single peak in the specific heat.

When $\delta > 1$ we enter the ferromagnetic ground state at low temperatures and observe a sharp peak in the specific heat at the critical temperature $T_c$, consistent with a conventional symmetry-breaking phase transition. 
The critical temperature as a function of $\delta$ can be extracted from the position of the specific heat peak.
{The heat capacity shows more structure in addition to the sharp peak found at the onset of long-ranged ferromagnetic order. 
Using the $\delta=1.2$ curve as an illustrative example, there are three peaks: a broad low-temperature Schottky peak as type 0 tetrahedra freeze-out, a sharp central peak at $T_c$, and another broad Schottky peak at high temperature as type 1A tetrahedra freeze-out. 
For larger $\delta$ the two Schottky peaks are not independently resolved, but still visible as broadening around the sharp peak.}
{In the thermodynamic limit, we expect the sharp central peak in the specific heat to diverge, while the broad features are system size-independent. See App.~\ref{app:further_numerical} for further discussion.}

\paragraph*{Monopole density.}
The single monopole number density, $n$, is a clear indicator of the phase realized at a certain temperature and distortion $\delta$.
Here, we only show the total number of single monopoles, combining types 1A and 1B but not counting the double monopoles.
At high temperatures, within the paramagnetic phase, all states are equally likely, yielding $n=8/16=1/2$.
In the spin ice phase $-1 < \delta < 1$, shown in panels (d,e) of \autoref{fig:thermo}, the density drops to zero as ground states are ice configurations without any monopoles.
At the phase boundaries, $\delta=\pm 1$, type 0 and type 1B/1A configurations are degenerate, and the density can be well understood by simple state counting.
Remarkably, for $\delta=-1$, the darkest red curve in (d), $n$, remains constant at $1/2$. This is due to the degeneracy of the six type 1A and the six type 0 configurations and, in addition, the degeneracy of type 2 and the type 1B configurations. The freezing out of excitations, therefore, leaves no trace in $n$ despite clearly manifesting in the specific heat and entropy. 
At $\delta =1$, the lightest blue curve in (f), $n$ drops to one quarter as the two type 1B and the six type 0 configurations are degenerate: $n=2/(2+6)=1/4$.
In the ferromagnetic regime, for $\delta >1$, the density approaches $n=1$.
The {total} monopole density initially drops as the type 1A configurations are {converted into type 0 and type 1B configurations}. The density then begins to increase at temperatures slightly above $T_c$ as the system arranges itself into a type 1B monopole crystal.

\paragraph*{Residual entropy.}
{Spin ice, in all its incarnations,} is famously known for its residual entropy.
Panels (g,h) in \autoref{fig:thermo} show the entropy obtained by integrating the specific heat from the high-temperature limit according to 
\begin{equation}
    S(T) = S_\infty - \int_T^\infty {\rm d}T' \frac{C_V(T')}{T'}
    \, ,
\end{equation}
where $C_V(T)$ is the specific heat and $S_\infty = N \log(2)$ is the infinite temperature entropy for a system of Ising spins. 

The number of spin ice ground states grows exponentially with the volume, and the system is said to have an extensively degenerate set of ground states for $-3\leq\delta\leq 1$.
As pointed out in \autoref{sec:GS and excitations}, the total number is known to be accurately estimated using a method originally devised by Linus Pauling to compute the residual zero-temperature entropy of water ice \cite{pauling1935}.
A generalized Pauling entropy estimate can be structured as follows: For every tetrahedron, there are sixteen possible spin configurations, of which $k$ minimize the energy on that tetrahedron.
If we treat the state of each tetrahedron as an independent constraint, a system with $N$ spins ($N/2$ tetrahedra) has approximately $2^N \left(k/16 \right)^{N/2} = \left(k / 4 \right)^{N / 2}$ ground states.
This results in a residual entropy of 
\begin{equation}
    S = \frac{k_B N }{2} \ln{\left(\frac{k}{4}\right)}
    \, .
    \label{eq:entropy}
\end{equation}

For $|\delta| < 1$, $k=6$ and the residual entropy per spin is $s = S / (k_bN) = \frac{1}{2}\ln{\left(3 / 2 \right)}$.
This is also true for $-3 < \delta < -1$ since each ice ground state can be mapped to a valid monopole configuration using the transformation in \autoref{eq:transformation}.
At $\delta=1$, the six two-in-two-out configurations have the same energy as the two type 1B monopole configurations.
Similarly, at $\delta=-3$, the six type 1A monopole configurations have the same energy as the {all-in} and {all-out} (type 2) configurations.
Thus, for these two points $k=8$ and $s = \frac{k_B}{2}\ln{\left(2 \right)}$ as illustrated by the lightest blue curve in (h).
The residual entropy is even higher at the duality point, $\delta=-1$, with $k=12$ as the six type 0 and six type 1A configurations are all degenerate.
This gives the estimate $s = \frac{k_B}{2}\ln{\left(3 \right)}$, consistent with the Monte Carlo result highlighted in dark red in (g).

The ordered phases at $\delta < -3$ and $\delta > 1$ have $s = 0$. 
Here, it is sufficient to pick the direction of a single spin to fix all other spins in the system.
The constraints from different tetrahedra can no longer be treated as independent, and the Pauling estimate does not apply.

\paragraph*{Magnetic susceptibility.}
The susceptibility becomes anisotropic since the distortion in our model breaks the lattice symmetry. Here, we only show the magnetic susceptibility in the $[111]$ direction for $\delta > 1$, panel (i) of \autoref{fig:thermo}.
It develops sharp peaks at the same temperatures as the specific heat and is a clear sign of a symmetry-breaking phase transition. An analysis of the critical exponents of the susceptibility and specific heat close to $T_c$ is provided in \appref{app:ordering}. 
{We also plot the susceptibility at $\delta=1$ (light blue) for comparison, which increases as $\propto T^{-1}$ upon cooling.}

\section{Dynamics \label{sec:dynamics}}
The preservation of the extensive ice degeneracy for a broad range of strain in our model, in contrast with previous studies \cite{jaubert2010spin, xie2015magnetic}{, where strain along the $[100]$ crystallographic axis was considered}, makes it possible to apply the standard model of spin ice dynamics \cite{jaubert2009, Jaubert2011}. There, the dynamics of the low-energy excitations is modeled under the assumption that the system evolves through random, independent single-spin flips occurring at a characteristic timescale $\tau_0$, \cf \appref{app:numericalmethods} for details. 
More precisely, we perform heat-bath Monte Carlo simulations with uniform single-spin flip attempt rates. 
This {is the simplest form of dynamics one can imagine}, and therefore allows us to focus solely on the impact of energetic constraints imposed by strain{, as opposed to any dynamical constraints imposed by spin flipping rules}. Qualitatively similar constraints on monopole motion arise in the ``beyond the standard model''  for spin ice dynamics \cite{hallen2022}, albeit from a different origin as dynamical rules on spin flip rates come from the quantum mechanical description rather than energetic barriers imposed by strain. {The beyond the standard model is commented on further in \autoref{sec:experiments}, but will not be used in this work.}

\subsection{Phenomenology \label{subsec:pheno}}
\begin{figure}[t]
    \includegraphics[width=\columnwidth]{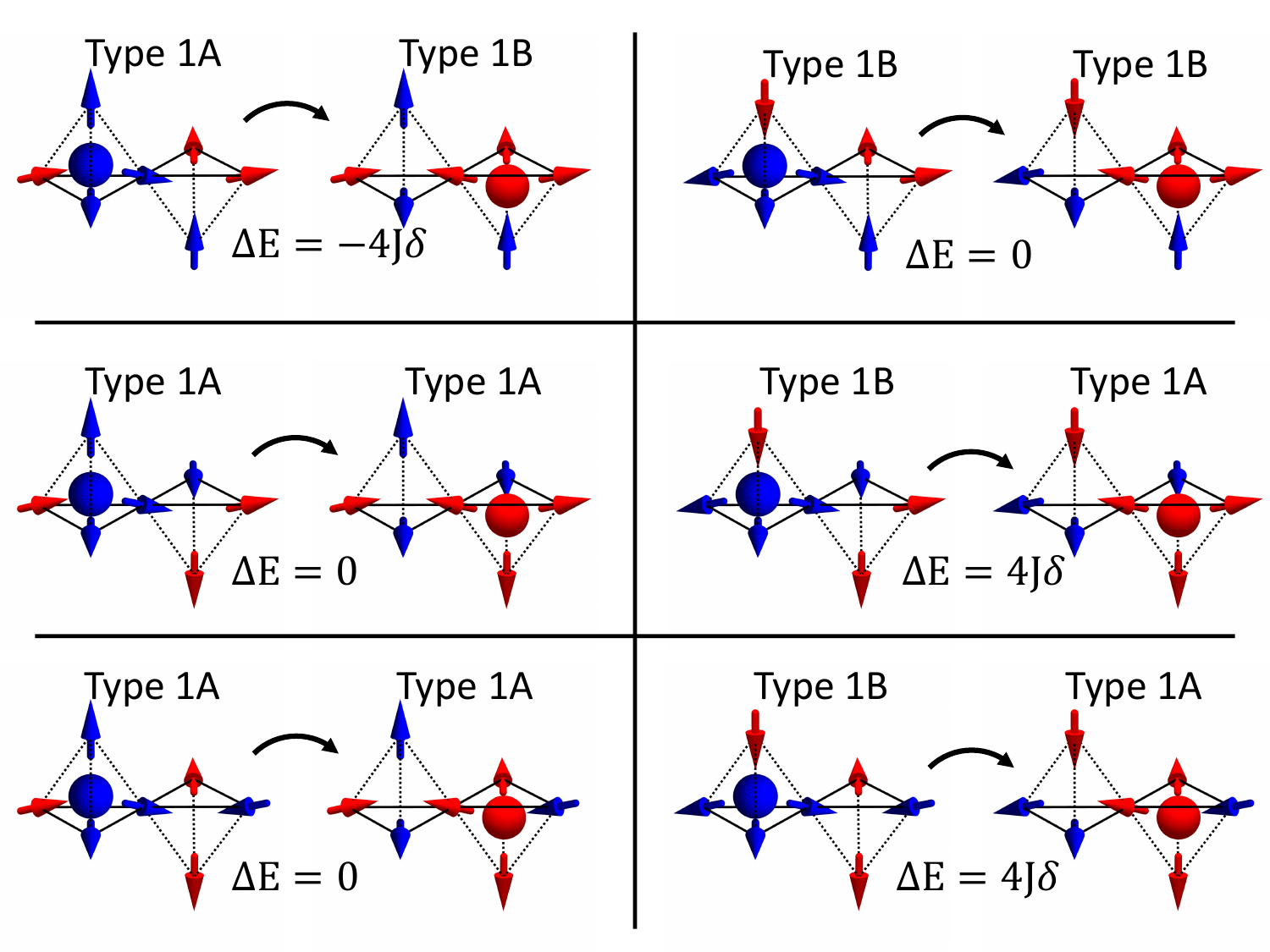}
    \caption{Possible moves for monopoles of type 1A (left) and type 1B (right) with the associated energy differences. While type 1A monopoles have their minority spin within the $(111)$ plane, the minority spin of type 1B is a member of $\SL$. Whether a move that converts the monopole from type 1A to type 1B is energetically favorable depends on the sign of $\delta$. Moves through the spin on $\SL$ are not shown; they are always allowed for type 1A monopoles and never allowed for type 1B monopoles.}
    \label{fig:allowed_moves}
\end{figure}

We focus on the magnetic monopole excitations in the spin ice phase for $\vert\delta\vert < 1$.
The lowest-energy excitations are single monopoles with an energy cost $J(2-2\vert\delta\vert)$. 
A monopole moves when one of the three majority spins in its tetrahedron flips, causing the monopole to hop to the neighboring tetrahedra, see \autoref{fig:allowed_moves}. 
Flipping the minority spin instead generates a new monopole pair on top of the existing monopole. The energy penalty associated with this process is large, and such events are effectively forbidden.

Depending on the sign of $\delta$, either type 1A or type 1B monopoles are lower in energy -- resulting in either type A or type B dynamics as indicated in \autoref{fig:PD_strain}. The energy difference between the two monopole types is $4 J \vert \delta \vert$. 
Hence, at high temperatures $T \gg 4J\vert \delta \vert$, the monopoles can readily change types, and the monopoles are isotropically diffusive as in the isotropic case without strain \cite{ryzhkin2005,jaubert2009,Jaubert2011,kirschner2018,hallen2022}. 
Here, we emphasize a temperature regime $T/J \ll 4\vert\delta\vert,\ 2-2\vert\delta\vert$, where the dynamics are dominated by the motion of a dilute set of ``cheap'' monopoles, with rare promotions to ``expensive'' ones. 
Cheap and expensive refer to the energies, $J(2-2\vert\delta\vert)$ and $J(2+2\vert\delta\vert)$, of either type of monopole configuration compared to a two-in-two-out configuration. 
The promotion occurs on a timescale 
\begin{align}
    \kappa_0 = \exp\left( J \frac{4\vert\delta \vert}{T}\right) \tau_0
    \, ,
    \label{eq:kappa0}
\end{align}
coming from the Boltzmann weight of the heat bath.
Moves that conserve or reduce the energy occur on a timescale $\tau_0$.

In the following, we focus on timescales smaller than $\kappa_0$ and discuss the possible zero-energy moves. Monopole motion is diffusive for $\delta=0$~\cite{hallen2022}, since the three majority spins are randomly distributed. 
This situation is in stark contrast to our anisotropic model, where energetic constraints are encoded in the rules governing if a monopole belongs to type 1A or 1B, leading to different levels of restrictions on monopole motion for the two types. Diffusive behavior is recovered in all cases at timescales larger than $\kappa_0$, as this is the typical timescale on which monopoles change type. 
{On this timescale lower-energy monopoles thus overcome energy barriers associated with changing type, which opens up further paths for them resulting in diffusive monopole motion.} 

\paragraph*{Type 1B dynamics.}
Type 1B monopoles are the lowest-energy excitations for $\delta > 0$ and have their minority spin on sublattice $\SL$. They cannot move with zero energy moves through sublattice $\SL$ spins, leaving them energetically constrained to move in planes normal to $[111]$. Furthermore, moves in the plane at zero energy cost are only possible if the monopole remains type 1B after the move, which requires the new minority spin to also be a sublattice $\SL$ spin. This is only true for one of the three possible configurations of the surrounding tetrahedra in the $(111)$ plane, \cf \autoref{fig:allowed_moves}. Therefore, type 1B monopoles can only move through, on average, one out of the three $\SLa$ spins at each tetrahedron. 

The centers of tetrahedra form honeycomb lattices in the $(111)$ planes.
Hence, we can understand the zero-energy motion of type 1B monopoles as a random walk on these honeycomb lattices with $2/3$ of the bonds removed at random. This can be viewed as a bond percolation model with an effective filling fraction $p_{(111)}=1/3$. However, the bond percolation threshold for the honeycomb lattice is $p_c \approx 0.65$ \cite{sykes1964exact}, and we thus expect monopoles to be localized to small clusters for $\delta > 0$ and $T \ll 4J \delta$.

One effect of the localization of type 1B monopoles is that the energetically more expensive type 1A monopoles play a significant role in the system relaxation. This introduces another characteristic timescale in the problem 
 \begin{align}
    \kappa_1 = \exp\left( J \frac{(2+2\vert\delta\vert)}{T}\right) \tau_0
    \, ,
    \label{eq:kappa1}
\end{align}
proportional to the density of type 1A monopoles in the $\delta > 0$ regime.

%
%
%
%
\paragraph*{Type 1A dynamics.}
For $\delta < 0$, type 1A monopoles are lower in energy. The minority spin of type 1A monopoles is \textit{not} on sublattice $\SL$, making their motion less constrained than that of type 1B monopoles. 
A majority spin can be flipped without energy cost if the sublattice $\SL$ spin in the new tetrahedron hosting the monopole is a majority spin after the flip.
In this case, motion through a sublattice $\SL$ spin, can \emph{always} happen at zero energy when there is not already a monopole in the adjacent tetrahedron.
Motion through each of the other two majority spins costs zero energy in $2/3$ of cases, \cf \autoref{fig:allowed_moves}. 
As $2/3$ of the $\SLa$ spins are majority spins, the probability that the monopole can move through an arbitrary spin on sublattice $\SLa$ is thus $p_{(111)}=4/9$. This is still much below the percolation threshold of the honeycomb lattice.
However, monopoles can propagate perpendicular to the honeycomb layer {through the $\SL$ spins}. 
This allows them to jump between finite clusters in the $(111)$ plane and leaves the monopoles diffusive with an anisotropic diffusion tensor favoring motion in the $[111]$ direction. 
In the language of (anisotropic) bond percolation, we can view their motion as a random walk on a diamond lattice with two filling fractions: $p_{[111]}=1$ for bonds along the $[111]$ direction and $p_{(111)}=4/9$ for the three remaining bond directions. 

{
We note that monopoles can, in principle, become dynamically localized in this regime. This occurs when all six moves taking a monopole away from two tetrahedra connected through the $\SL$ spin are blocked. The monopole is then only able to move between these two tetrahedra. Instances like these are rare, and we find that the vast majority of type 1A monopoles are not dynamically localized.
}

\paragraph*{Inverted spin ice.}
The same arguments apply to the inverted spin ice phase.
For $-3 < \delta < -1$, the ground states have type 1A configurations everywhere, \cf red part in \autoref{fig:PD_strain}.
The picture explained above is, however, directly applicable if one invokes the duality around $\delta=-1$.
The lowest-energy excitations in the regime $-3 < \delta < -2$ are type 2 tetrahedra (double monopoles), that move on top of a background of single monopoles with the same rules underpinning their motion as that of the type 1B monopoles.
In the regime $-2 < \delta < -1$, the lowest-energy excitations are instead type 0 tetrahedra that move on top of the background of single monopoles following the rules governing type 1A monopole moves.

\paragraph*{Ordered phases.}
The long-range ordered phases for $\delta < -3$ and $\delta > 1$ do not have point-like excitations.
Instead, excitations take the form of domain boundaries that do not move without an energy penalty.


\subsection{Mean-squared displacement}
\begin{figure*}
    \centering
    \includegraphics{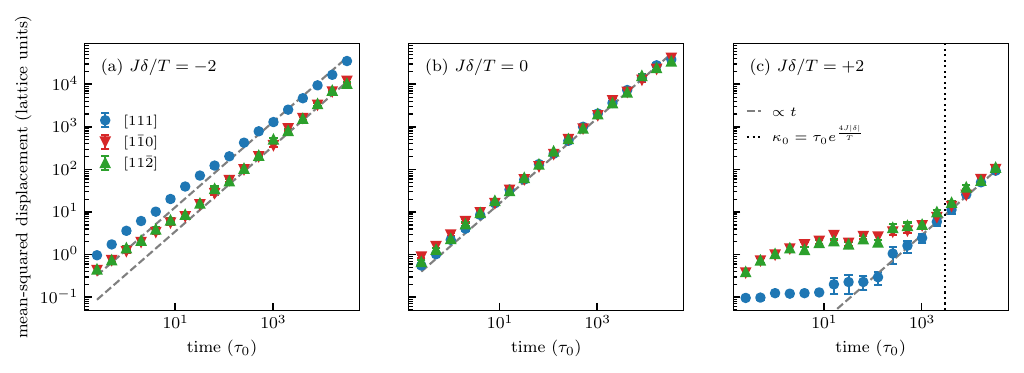}
    \caption{The mean-squared displacement (MSD) of individual isolated monopoles projected along the $[111]$ direction (blue circles) and along two orthogonal directions: $[1\bar{1}0]$ (red down triangles) and $[11\Bar{2}]$ (green up triangles). The MSD is measured in units of the squared distance between the center of two neighboring tetrahedra. 
    For $\delta < 0$, panel (a), type 1A monopoles are favored, and the monopole dynamics are essentially diffusive, albeit suppressed in the plane normal to $[111]$. The gray lines indicate diffusive behavior, $\propto t$, which is isotropic for $\delta = 0$ in panel (b). In stark contrast, localization is observed for $\delta > 0$, panel (c), where type 1B monopoles are favored. Here, the monopoles are restricted to small clusters on timescales smaller than $\kappa_0/\tau_0\approx 3\cdot 10^3$ (dotted black lines) from \autoref{eq:kappa0}. The localization is followed by diffusive behavior at longer times mediated by the thermal promotion of type 1B monopoles to higher-energy diffusive type 1A monopoles.}
    \label{fig:MSD}
\end{figure*}

The impact of strain on the monopole motion can be observed by tracking the trajectories of individual monopoles in Monte Carlo simulations. More details are provided in \autoref{app:numericalmethods}. \autoref{fig:MSD} shows the mean-squared displacement (MSD) of a single, isolated monopole moving without annihilation or creation events. This mimics the natural motion of monopoles at low monopole densities. To capture the anisotropic response of our model, the monopole displacement is projected along the $[111]$ direction, as well as along two orthogonal directions. The choice of these two orthogonal directions is arbitrary and has no impact on the observed MSD. Since creation and annihilation events are banned, the energetics are completely set by the ratio $J \delta /T$.

%
%
%
%

Consistent with our phenomenological arguments, anisotropic diffusive monopole motion in the MSD is observed in the type A regime ($\delta \leq 0$; panel (a) of \autoref{fig:MSD}). Motion in the $[111]$ direction is favored as expected. The response in the type B regime ($\delta > 0$; panel (c) of \autoref{fig:MSD}) is quite distinct. Here, the MSD initially appears to plateau, indicating that monopoles are localized to small clusters. This is consistent with type 1B monopoles never moving through $\SL$ spins, effectively restricting their motion to a small number of tetrahedra within the $(111)$ plane. Indeed, the localization is stronger in the $[111]$ direction\footnote{Note that even in the limit of $J \vert\delta\vert /T\to \infty$, the MSD projected along $[111]$ will not be zero, as monopole moves through any spins in $\SLa$ have a nonzero component along $[111]$.}. While type 1B monopoles are initially localized, promotion events occur on a timescale $\kappa_0$, and this process restores diffusive monopole motion. 
In contrast to the type A regime where movement along $[111]$ is favored, the diffusion is isotropic on timescales of $\kappa_0$. Once an expensive type 1A monopole is created, its motion is not restricted in any direction because its propagation never enhances the energy. Therefore, as long as no double monopole is created, every move is accepted.

\paragraph*{Subdimensional diffusion.}
Within a spin ice state, any closed loop of aligned spins can be flipped at zero energy cost. The smallest such loop consists of six spins forming a hexagonal plaquette. These ``ring exchange'' moves naturally arise within the context of \emph{quantum spin ice} as the lowest-order relevant perturbation to the model, leading to quantum fluctuations mixing the classically degenerate spin ice states \cite{hermele_2004}. 

\begin{figure*}[t]
    \centering\includegraphics[width=1.\linewidth]{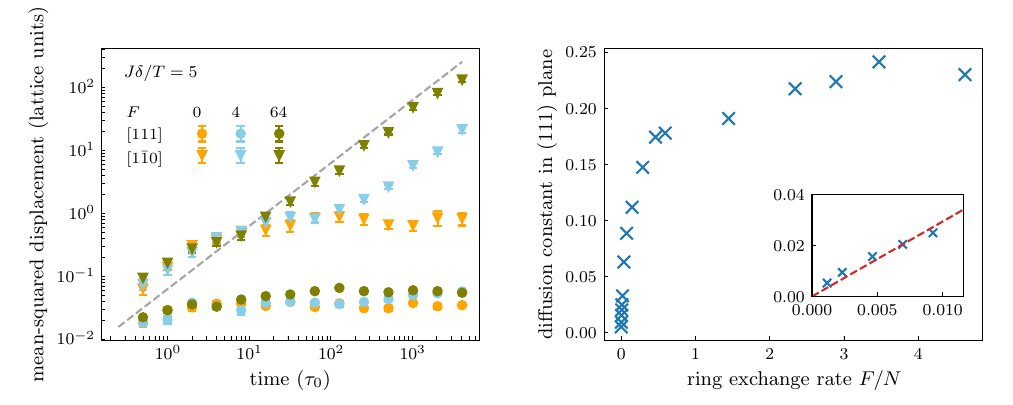}
    \caption{\textit{Left}: Mean-squared displacement for a single type 1B monopole in the presence of ring exchange updates that cause zero-energy fluctuation within the ice background.
    The rate $F$ of ring exchange updates to attempted single-spin flips are given in the legend. Note that this is the rate of \emph{performed} updates, not the attempt rate. Here, we use moderate ring exchange rates to observe subdimensional diffusion at relatively short times. A smaller $F$ leads to subdimensional diffusion at late times, and delocalization due to ring exchange would then compete with delocalization due to the thermal promotion to type 1A monopoles. \textit{Right}: Subdimensional diffusion constant plotted versus the ring exchange rate $F$ normalized by the number of spins $N$. 
    The diffusion constant is measured in units of $a_d^2/\tau_0$, where $a_d$ is the distance between the center of two neighboring tetrahedra, and was extracted from linear fits to the MSD in the $[1\bar{1}0]$ direction. 
    The subdimensional diffusion constant shows expected linear dependence with $F/N$ in the $F \rightarrow0$ ($\tau_0 \ll \tau_{lock}$) limit, highlighted in the inset, and the transition to a plateau region at large $F/N$, corresponding to the $\tau_0 \gg \tau_{lock}$ limit.
    }
    \label{fig:MSD+Loops}
\end{figure*}

The classical analog of the ring exchange moves consists of stochastically finding a hexagonal plaquette with aligned spins and flipping the six spins. In our model, this lifts the localization of type 1B monopoles, as shown in \autoref{fig:MSD+Loops}. The localization is only lifted in the (111) plane as monopoles remain localized in the [111] direction. This is because any ring exchange update, including a $\SL$ spin neighboring a monopole, converts the monopole from type 1B to type 1A at an energy cost of $4 J \delta$. This process is thermally suppressed for $T \ll 4 J \delta$. 
Thus, when ring exchange updates are included, the monopoles exhibit \emph{subdimensional diffusion} at zero temperature -- reminiscent of fractonic motion \cite{pretko2020fracton, nandkishore2019fractons}. We also note that flipping an open string of four spins attached to a type 1B monopole can move it along the $[111]$ direction while preserving the energy.

%
%
%
%

The subdimensional diffusion of the monopole can be understood in terms of a random walk on a honeycomb lattice where the traversability of a bond changes in time. 
In a completely unconstrained random walk on a honeycomb lattice, meaning that monopoles can move freely in the (111) plane, diffusion can be observed with a constant $D$: $\langle r^2\rangle = D t$.
However, for localized 1B monopoles, moves across any bond are only possible in one of three cases.
The additional ring exchange updates scramble the ice background and allow new moves across previously disallowed bonds.
The typical timescale, $\taulock$, it takes to unlock such a bond scales inversely proportional to the ring exchange rate $F$.
Given the single-spin flipping rate $\tau_0$, there are two important limits: (i) $\tau_0\gg \taulock$ and (ii) $\tau_0\ll \taulock$.
For limit (i), where the ice background is fully scrambled between two attempted hops, the probability of any attempted hop being accepted is $1/3$, and this effectively reduces the diffusion constant to $D/3$. 
In contrast, for limit (ii), the time a monopole is localized to a specific cluster is governed by $\taulock$. One can model this as monopoles performing a random walk with a new hopping time $\taulock$, leading to a reduced diffusion constant $D (\tau_0 / \taulock) \propto D F$. In this limit, the diffusion constant thus scales linearly with the ring exchange rate $F$.

\subsection{Magnetic noise}\label{sec:psd}
The stochastic thermal motion of monopoles generates magnetic noise, which is experimentally observable and has been measured in spin ice materials \cite{dusad2019magnetic, samarakoon2022anomalous, dasini2024discovery, morineau2024satisfaction}. 
Within the spin ice regime $-1 < \delta < 1$, we characterize the thermal magnetic noise of our model by extracting the power spectral density (PSD) for the different spin subspecies. The subspecies PSD is defined as
\begin{equation}
    \operatorname{PSD_{\cal S}}(\omega)=\frac{1}{N_{\cal S}}\left\langle|\tilde{M_{\cal S}}(\omega)|^2\right\rangle
    \, ,
    \label{eq:psd}
\end{equation}
where $N_{\cal S } = N/4$ is the number of spins that belong to subspecies ${\cal S}$, and $\tilde{M_{\cal S}}(\omega)$ is the Fourier transform of the temporal trace of their total magnetization. 
For diffusive monopole motion, the PSD is a Lorentzian with functional form $\propto (1 + \omega^2 \tau^2 )^{-1}$, where $\tau$ is the relaxation time \cite{hallen2022}. Characteristic for a Lorentzian is the low-frequency plateau for $\omega \lesssim 1/\tau$ and $\omega^{-2}$ decay for $\omega \gtrsim 1/\tau$. 

%
%
\begin{figure*}[htbp]
    \centering
    \includegraphics[width=1.\textwidth]{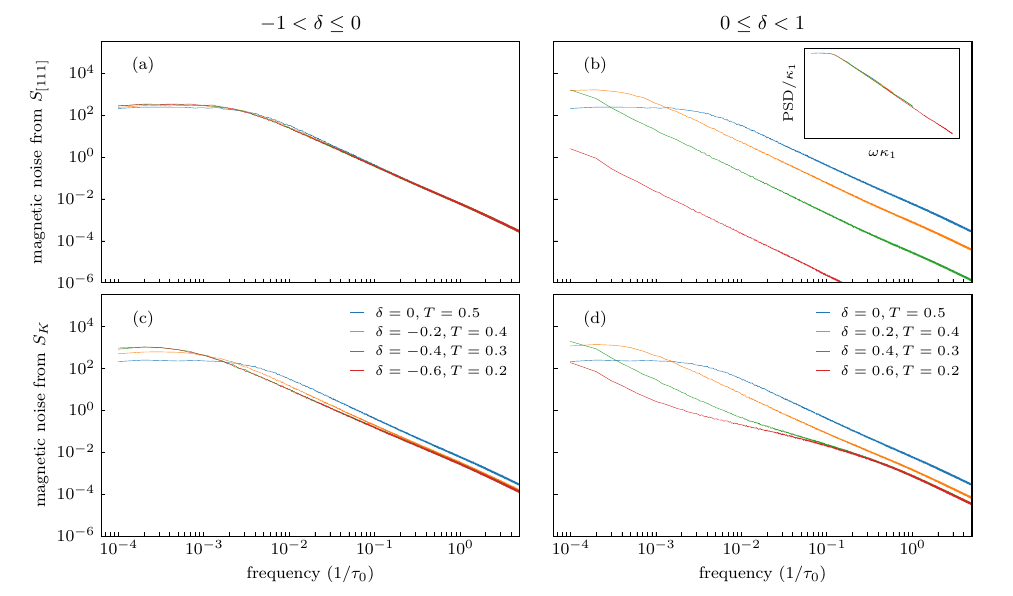}
	\caption{Power spectral density (PSD) of the magnetic noise, \cf \autoref{eq:psd}, measured from spin subspecies $\SL$ (top row) and $\SLa$ (bottom row). The frequency is in units of the spin-flip rate, which is, \eg, for Dy$_2$Ti$_2$O$_7$ approximately $10^4\ {\rm s}^{-1}$. We show data for the different spin ice regimes $-1<\delta \le 0$ and $0\le \delta < 1$ in the left and right columns, respectively. The values for $\delta$ and $T$ are chosen such that a finite but small number of cheap monopoles are present in our simulation of $N=4000$ spins. For $\delta < 0$, type 1A monopoles are energetically favored and propagate diffusively, yielding a Lorentzian noise spectrum for both subspecies in the left column, panel (a,c). Motion of type 1B monopoles, which are favored for $\delta > 0$, are much more constrained and are displayed in the right column. Since type 1B monopoles cannot propagate through $\SL$ spins, the Lorentzian observed in panel (b) is attributed to the creation, annihilation, and propagation of expensive monopoles. Thus, on times scales of expensive type 1A monopoles, $\kappa_1$ from \autoref{eq:kappa1}, we find a perfect scaling collapse in the inset when considering rescaled units. The anisotropy of our model manifests itself most prominently in panel (d), revealing a suppressed noise for intermediate frequencies coming from the localization of cheap monopoles.}
    \label{fig:PSD}
\end{figure*}
%
%
\autoref{fig:PSD} shows the PSD for subspecies $\SL$ in the top row and the PSD of \textit{one} different subspecies, $\SLa$, in the bottom row. We fix the temperature such that $(2 - 2\vert\delta\vert) J/ T = 4$, ensuring that the average density of the low-energy monopoles is the same for different values of $\delta$. These simulations are initialized from thermal equilibrium and we now allow for monopole creation and annihilation.

\paragraph*{Type 1A dynamics.}
Focusing on type 1A dynamics first ($-1 < \delta \leq 0$; left column of \autoref{fig:PSD}), we find that the PSDs of all spin subspecies are essentially Lorentzian, consistent with the predicted anisotropic diffusive monopole motion observed in panel (a) of \autoref{fig:MSD}. The PSD of subspecies $\SL$ depends only weakly on $\delta$, which emphasizes the fact that type 1A monopoles are always free to move {through the $\SL$ spins}. 
The PSD of sublattice $\SLa$ displays a stronger $\delta$-dependence, and for increasing $\delta$, the low-frequency plateau appears to shift to lower frequency and larger amplitude, consistent with an increase in the relaxation time $\tau$. This is a manifestation of the increased energy cost associated with converting a type 1A to a type 1B monopole, which reduces the monopole mobility through subspecies $\SLa$.

\paragraph*{Type 1B dynamics.}
The monopole motion is strongly constrained for $\delta > 0$, and this manifests itself in the PSD ($0\leq \delta<1$; right column of \autoref{fig:PSD}). 
The fluctuations of spins belonging to $\SL$, panel (b), come purely from the motion, creation, or annihilation of type 1A monopoles, which are now energetically more expensive than type 1B. 
The resulting PSD is Lorentzian, as in the $\delta < 0$ case, but with significantly suppressed magnitude and an increased relaxation time owing to the lower density of type 1A monopoles in this regime. 
Rescaling the PSD and frequency by $\kappa_1$ from \autoref{eq:kappa1}, proportional to the density to type 1A monopoles, we find a perfect scaling collapse as illustrated in the inset of panel (b) in \autoref{fig:PSD}; see also \autoref{ssec:scaling} for further details. 

The PSD for spins belonging to $\SLa$, panel (d) of \autoref{fig:PSD}, behaves quite differently in the $\delta>0$ regime. Here, the high-frequency noise is primarily generated by the localized motion of type 1B monopoles.
For $\omega \gtrsim 1/\tau_0$ the PSD displays a $\omega^{-2}$ decay, which is typical for PSD curves measured in Monte Carlo simulations \cite{samarakoon2022anomalous}.
For larger values of $\delta$, there is a regime of intermediate frequencies where the PSD decays significantly slower with frequency -- a manifestation of the monopole localization.
This slower decay would turn into a plateau persisting to arbitrarily small frequencies as $T \to 0$ if the only relaxation mechanism in the system was the zero-energy motion of type 1B monopoles. However, both the presence of a small but nonzero density of type 1A monopoles and the spontaneous creation of new type 1B monopoles offer alternate avenues for the system to relax. This manifests as a faster decay of the noise spectrum at low frequencies. Below some frequency, the noise must plateau in all cases (unless $T\to 0$) as all spins can flip at a finite energy cost, and beyond some timescale, the system is therefore completely uncorrelated from its previous configurations. This frequency is, however, not accessible in our simulations for $\delta > 0$.

\begin{figure*}
    \centering
    \includegraphics[width=\textwidth]{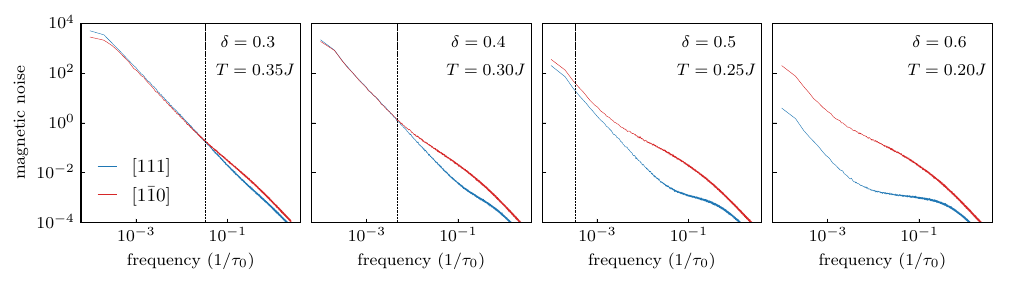}
    \caption{PSD of the magnetization measured along the $[111]$ and $[1\Bar{1}0]$ direction.  Note how the distortion leads to an anisotropic response, with a suppression of the noise along the $[111]$ direction as compared to the orthogonal $[1\bar{1}0]$ direction. The vertical dashed black lines indicate the inverse of $\kappa_0 = {\rm e}^{4 J \delta / T}\tau_0 $, which is the approximate timescale on which the type 1B monopoles are converted to higher energy type 1A monopoles. It clearly separates the short-time anisotropic dynamics mediated by zero-energy moves from the long-time isotropic behavior. }
    \label{fig:fig0_PSD}
\end{figure*}

Experimentally, there has recently been renewed interest in the dynamics of magnetic monopoles~\cite{dusad2019magnetic, Nisoli2021, samarakoon2022anomalous, hallen2022, hsu2024dichotomous}, largely driven by the ability to probe magnetic noise spectra using SQUIDs.
The magnetic noise plots in \autoref{sec:dynamics} took advantage of the microscopic nature of the Monte Carlo simulation to present sublattice-resolved PSDs. For completeness, we present data for the (experimentally accessible) directional PSD curves in \autoref{fig:fig0_PSD}. 
These can be obtained from the sublattice resolved data in \autoref{fig:PSD}.
We choose moderate values of $\delta/T$ that might be realized in future experiments using the sublattice doping discussed in the next section. As expected, we find stark differences {between magnetic noise measured along the $[111]$ direction and along its orthogonal direction $[1\Bar{1}0]$.} \autoref{fig:fig0_PSD} also demonstrates the transition between short-time anisotropic and long-time isotropic behavior. This crossover occurs on the timescale $\sim \kappa_0$, beyond which the distinction between the monopole types loses its meaning. 

\section{Experimental realizations \label{sec:experiments}}
Famously, spin ice is approximately realized in nature by the rare-earth pyrochlore compounds Dy$_2$Ti$_2$O$_7$~\cite{ramirez_zero-point_1999} and Ho$_2$Ti$_2$O$_7$~\cite{harris_geometrical_1997,bramwell_frustration_1998}.
Experiments at thermodynamic equilibrium are possible down to temperatures of approximately 0.55~K \cite{snyder2004low}, revealing characteristic properties such as pinch-points~\cite{morris_dirac_2009,fennell_magnetic_2009} as well as signatures of the magnetic monopoles~\cite{Castelnovo2008}. 0.55~K corresponds to roughly $T/J\approx 0.25$ in our nearest-neighbor model \cite{bramwell2020history}.

There are different approaches to engineer the anisotropic model, \autoref{eq:Hamiltonian} with $\delta\neq 0$.
The simplest way to achieve this is the application of uniaxial stress along the $[111]$-axis.
Naively, this reduces the distances between atoms in the direction of the applied stress.
A rough estimate, neglecting the influence of the oxygen environments or spin alignment~\cite{edberg2020effects}, is based on the dipolar interaction that predicts an increase of the coupling constant scaling with $r^{-3}$.
For rare-earth pyrochlores, hydrostatic pressure reaching values up to $\sim$50~GPa has been achieved~\cite{rittman_strain_2017,suganya_structural_2020}, which reduces, for example, the lattice constant in In$_2$Mn$_2$O$_7$~\cite{li_structural_2021} from \SI{9.7113}{\angstrom} at ambient pressure to \SI{9.3764}{\angstrom} at a pressure of $29\,$GPa.
In the naive dipolar approximation, the nearest-neighbor interaction strength would increase by $\sim$10\% under these conditions.

Uniaxial stress is more limited than hydrostatic pressure, with previous experiments demonstrating stresses of up to $2.2\,$GPa ($1.3\,$GPa) for Ho$_2$Ti$_2$O$_7$ (Dy$_2$Ti$_2$O$_7$)~\cite{mito2007uniaxial,edberg2019dipolar,edberg2020effects,pili2022topological}. 
By modeling the magnetization with respect to the applied stress and an external field, Refs.~\cite{edberg2019dipolar,edberg2020effects} estimate the increase in the nearest-neighbor coupling parallel to the stress axis and the lattice compression.
Taking both dipolar and exchange interactions, with the dipolar spin ice Hamiltonian~\cite{den_hertog_dipolar_2000}, into account, we expect this to correspond to an increase of the effective nearest-neighbor coupling of $\delta\sim 0.06$ and $\delta\sim 0.02$ for Ho$_2$Ti$_2$O$_7$ and Dy$_2$Ti$_2$O$_7$, respectively.

The novel monopole behaviors described in \autoref{sec:dynamics} are largely governed by the energy difference between type 1A and type 1B monopoles: $4\vert \delta \vert J$ in our model.
It is the ratio $4\vert \delta \vert J/T$ that determines the timescales on which anisotropic diffusion and monopole localization occur.
Combining the predicted values of $\delta$ from the experiments in Refs.~\cite{mito2007uniaxial,edberg2019dipolar,edberg2020effects} and the lowest accessible temperature, 0.55~K \cite{snyder2004low}, places us in a region where anisotropic effects are barely visible in simulations.
However, due to the cubic scaling of the dipolar interaction with distance and the exponential dependence of monopole motion on the energy-temperature ratio, small increases in the achieved strain have large effects on the dynamics. 
We are, therefore, optimistic that future high-pressure experiments will be able to observe parts of the phenomenology we have identified. 

%
%
%
%

Another mechanism to achieve larger magnitudes of $\vert\delta\vert$ involves substituting spins located on the sublattice $\SL$ with a species of spins characterized by a different magnetic moment.
This approach might be capable of exploring the full phase diagram in \autoref{fig:PD_strain}, which contains ferromagnetic and antiferromagnetic bonds.
For example, introducing spins with a larger or smaller magnetic moment{ into the $\SL$ positions of a spin compound leads to variations in $\delta$.}

Extending this idea, atoms that interact antiferromagnetically with the other species open up the possibility to observe phases predicted for the $\delta < -1$.
A notable example of this approach are tripod kagome systems where atoms on the sublattice $\SL$ are substituted with nonmagnetic ions~\cite{dun_magnetic_2016,paddison_emergent_2016}.
They realize the extremal case $\delta=-1$, where we have disconnected kagome planes.
Note, however, that doping atoms at the right position is extremely challenging, and it is much easier to place them randomly within the crystalline structure. 
Hence, understanding and modeling the Hamiltonian where $\delta$ is randomly assigned to a subset of spins is potentially important future work.

{We used standard model dynamics in \autoref{sec:dynamics}, which has proven to be an excellent model for the study of real materials \cite{Jaubert2011}. It should, however, be noted that more detailed and material-specific model dynamics have been developed~\cite{tomasello2019, hallen2022}. In particular, the beyond the standard model, which replaces the single global flipping attempt rate with two rates based on a microscopic quantum description of the system, can reproduce dynamical phenomena observed in Dy$_2$Ti$_2$O$_7$ (without strain) which cannot be reproduced using the standard model \cite{hallen2022, hsu2024dichotomous}. To what extent the different model dynamics apply to specific materials under strain (or with substitutional doping) remains to be ascertained and requires quantum mechanical microscopic considerations beyond the scope of this work.}

Lastly, a growing effort in artificial spin ice systems~\cite{skjaervo_advances_2020, may2021magnetic} and {thin films~\cite{jaubert_spin_2017,lantagne_spin_2018,Bovo2019,wen_epitaxial_2021} might provide an avenue to engineer strained models that potentially host similar localization physics.
In particular, thin films allow for a variety of different strain patterns that potentially lock monopoles within the layers.}

\section{Conclusions \label{sec:conclusions}}
Our simple model Hamiltonian $H_S$ in \autoref{eq:Hamiltonian}, nominally describing spin ice under uniaxial $[111]$-strain, has turned out to support a rich and varied physical landscape. Using the grouping of the sixteen single-tetrahedron configurations into four degenerate types, we have successfully described the full phase diagram of $H_S$. Remarkably, the extensive degeneracy of the ice ground states is preserved for a wide range of distortion values, which is in stark contrast to the behavior of uniaxial stress along other crystallographic axes~\cite{jaubert2010spin}. 
In contrast, the thermodynamic properties contain information about the distortion, for example, in the form of two peaks in the specific heat. 
The distortion further manifests itself in the dynamical properties, with two dynamical regimes appearing in the low-temperature limit. Using single-spin flip dynamics, we find that the magnetic monopoles split into a diffusive and a localized type, which dominate the dynamics at negative and positive $\delta$, respectively. 
For $\delta>0$, the localization of the low-energy monopoles suppresses the magnetic noise in a characteristic, anisotropic fashion. The timescale on which the monopoles remain localized grows exponentially with inverse temperature, making their localization a relevant property at low temperatures regardless of the strength of the distortion.
This, presumably, also leads to an exponential slow-down of thermalization at low temperatures.

Including additional dynamic processes, such as the ring exchange updates discussed here, partly restores the diffusive motion of the localized monopoles. In this case, the monopoles remain localized in the strain direction -- effectively making them subdimensional excitations.

Several indicators of novel monopole dynamics we identify are experimentally accessible~\cite{edberg2019dipolar, edberg2020effects, dusad2019magnetic, samarakoon2022anomalous, pili2022topological}.
This includes not only the magnetic noise spectra but also measurements of thermodynamic properties such as the specific heat. Furthermore, the impact of uniaxial strain on transport properties and out-of-equilibrium monopole dynamics, which are beyond the scope of this paper, is another interesting avenue for further research.
Although additional work is required to fully model the impact of strain on real spin ice materials, with the inclusion of long-range dipolar interactions as an obvious route for improvement, we believe the phenomenology identified here to be largely applicable. 
We hope our prediction will stimulate further experimental studies of spin ice under strain, focusing on the two distinct monopole types.

While we only considered global changes to the spin-spin interactions along the $[111]$ direction, other intriguing possibilities exist. One of these is the application of a strong uniaxial electric field. 
Another is the partial substitution of atoms on sublattice $\SL$, leading to a Hamiltonian where the distortion only appears on a fraction of the tetrahedra.
Finally, it has recently been proposed that monopoles in Dy$_2$Ti$_2$O$_7$ move on emergent dynamical fractals \cite{hallen2022, hsu2024dichotomous}. This opens up another interesting avenue for future work on the interplay of these fractals and energetic barriers induced by perturbations such as strain.

\section*{Acknowledgements}
We are thankful for stimulating discussions with Evan Smith, Benedikt Placke, Owen Benton, Roderich Moessner, and Claudio Castelnovo.
RS acknowledges the AFOSR Grant No. FA 9550-20-1-0235. 
CRL thanks the Max Planck Institute for the Physics of Complex Systems for its hospitality. 
This work was in part supported by the Deutsche Forschungsgemeinschaft under grants SFB 1143 (project-id 247310070) and the cluster of excellence ct.qmat (EXC 2147, project-id 390858490).

\bibliography{references}

\appendix


\section{Spin and magnetization convention
\label{app:spin_notation}}
{\paragraph{Spin convention}
The spins in our model are represented by classical vectors $\vec{s}$ that are constrained to point along the local $\langle 111 \rangle$ directions, i.e., a spin on sublattice $\alpha$ points along the unit vector $\vec{e}_\alpha$ with
\begin{eqnarray}
    \vec{e}_1 = \frac{1}{\sqrt{3}}
    \begin{pmatrix} 
    1 \\ 1 \\ 1
    \end{pmatrix} 
    &\qquad& 
    \vec{e}_2 = \frac{1}{\sqrt{3}}
    \begin{pmatrix} 
    1 \\ -1 \\ -1
    \end{pmatrix}  
    \\
    \nonumber
    \vec{e}_3 = \frac{1}{\sqrt{3}}
    \begin{pmatrix} 
    -1 \\ 1 \\ -1
    \end{pmatrix} 
    &\qquad& 
    \vec{e}_4 = \frac{1}{\sqrt{3}}
    \begin{pmatrix} 
    -1 \\ -1 \\ 1
    \end{pmatrix} \,,
\end{eqnarray}
where $\vec{e}_1$ points along the global $[111]$ direction and spins on sublattice $\SL$ are aligned with $\vec{e}_1$. Each spin can be expressed in terms of Ising variables as $\vec{s}_i = \sigma_i \vec{e}_\alpha$ where we take the spin length to be one. In our convention, $\sigma=+1$ corresponds to a spin pointing from an ``up'' tetrahedron to a ``down'' tetrahedron. With respect to \autoref{fig:pyrochlores}, this refers to pointing from a blue to a red tetrahedron.}

{
When rewriting our Hamiltonian, \autoref{eq:Hamiltonian_vector}, in terms of the Ising variables we make use of the fact that $\vec{s}_i \cdot \vec{s}_j = \sigma_i \sigma_j \vec{e}_\alpha \cdot \vec{e}_\beta = - \sigma_i \sigma_j / 3$ for any nearest-neighbor pair of spins.}

{\paragraph{Magnetization convention} Two different versions of the magnetization are used in this work. The directional magnetization is defined as
\begin{equation}
    M_{\vec{n}} = \vec{n} \cdot \sum_i \vec{s}_i 
    \, ,
\end{equation} 
for some unit vector $\vec{n}$.
The equilibrium $[111]$-directional susceptibility displayed in \autoref{fig:thermo} is given by 
\begin{equation}
    \chi_{[111]} = \frac{1}{N} \left. \frac{dM_{[111]}}{dH} \right\vert_{H=0} = \frac{1}{N} \frac{\left.\langle M_{[111]}^2\right\rangle - \left\langle M_{[111]} \right\rangle^2}{k_B T}\,.
\end{equation}
In contrast, the sublattice magnetization is defined as 
\begin{equation}
    M_{\cal S} = \sum_{i \in \mathcal{S}} \sigma_i
    \, .
\end{equation}
where the sum is over all sites on sublattice ${\cal S}$. 
The magnetic noise computed both from the directional and the sublattice magnetization are discussed in \autoref{sec:dynamics}.}

\section{Numerical methods}
\label{app:numericalmethods}
All numerical results presented in this work were acquired through Monte Carlo simulations using the Metropolis algorithm and its variants. The majority of simulations were carried out on a sample of linear size $L=10$ with $L\times L \times L =1000$ fundamental unit cells, corresponding to a total of $N=4000$ spins and $2000$ tetrahedra. Periodic boundary conditions were used in all cases. 

Worm updates were utilized to speed up equilibration at low temperatures in the spin ice phase. 
These updates involve identifying a chain of spins aligned head to tail, forming either a closed loop or terminating at a monopole at one end. Subsequently, an attempt is made to simultaneously flip the chain of spins. The acceptance of any update follows the Metropolis probability, denoted as $P = \min \left[1, \exp (-\Delta E / T) \right]$, where $\Delta E $ is the overall change in energy. Note that accepted worm updates neither create nor annihilate any monopoles nor do they change the number of double monopoles. The only contribution to $\Delta E$ for worm updates is thus the conversion of type 1A to type 1B monopoles or vice versa. The ring exchange updates discussed in \autoref{sec:dynamics} are equivalent to closed loop worm updates of length six. 

For thermodynamic measurements, we simulated the system by annealing from a high temperature, applying both worm updates and single-spin flip updates at each temperature step. 
As mentioned above, we only performed simulations for $\delta \geq -1$ and then invoke the duality given in \autoref{eq:transformation} to deduce the behavior for $\delta < -1$.

The equilibrium system dynamics were simulated using the standard model of spin ice dynamics~\cite{jaubert2009, Jaubert2011}: Metropolis algorithm Monte Carlo using only single spin flip updates and defining the flipping attempt time $\tau_0=1$ MC-sweep (1 MC-sweep corresponds to $N$ attempted spin flips). A combination of single-spin flip and worm updates are used to initialize the system at the desired temperature, and the dynamical measurements were only begun once the system was in thermal equilibrium. The PSD was computed from time traces of the magnetization using Welch's method~\cite{welch1967}. In an infinite system, the monopole density only goes to zero as temperature goes to zero. However, in a finite system, if the temperature is chosen too low, all monopoles can vanish, preventing any magnetization dynamics. Thus, when measuring the magnetization noise, we ensured that we were always at temperatures where the actual monopole number did not drop to zero at any point during the simulation.

Finally, the MSDs of individual monopoles were measured by first cooling a system of 54,000 spins until an ice state without any monopoles was reached. A single pair of low-energy monopoles is then created at an arbitrary position. One of these monopoles was kept fixed at its original position, whereas the other monopole was allowed to move. Each Monte Carlo update now consists of randomly selecting one of the four spins surrounding the mobile monopole and flipping this with probability $\min [1, \exp(-\Delta E / T)]$ if the selected spin is a majority spin. As creation and annihilation are not allowed, the monopole dynamics only depend on the ratio $J \delta/T$. We define four such updates to be equivalent to time $\tau_0$. This mimics the motion of a monopole under standard model dynamics in the limit of very low monopole density. To minimize the impact of the static monopole on the measurement results, the mobile monopole was allowed to move for approximately $10^6 \tau_0$ before its trajectory was measured. The MSD curves shown in \autoref{fig:MSD} and \autoref{fig:MSDsupp} are computed by performing a bootstrap average of the trajectories of 200 independent monopoles.

%
%
\begin{figure}[htp]
    \includegraphics[width=\columnwidth]{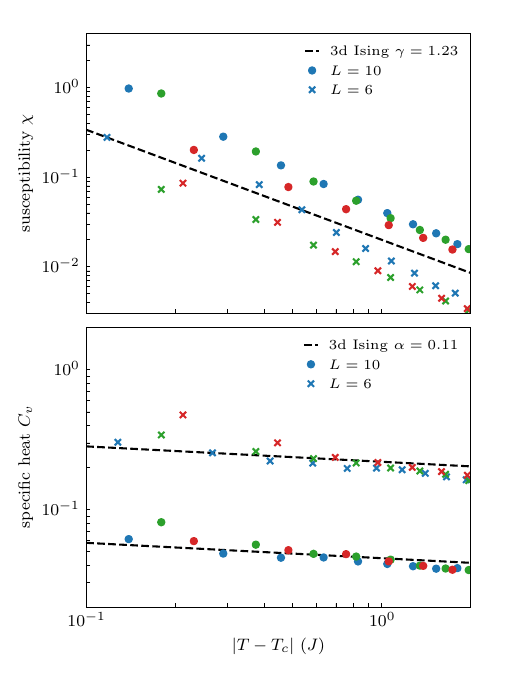}
    \caption{Susceptibility (top panel) and specific heat (bottom panel) plotted versus temperature for $\left|T- T_c \right|$. Distortion $\delta = 1.4,1.6,1.8$ (blue, green, and red) were considered, with two system sizes $L=6$ (cross) and $L=10$ (circle). The data appear to be approximately consistent with the 3d Ising critical exponents (dashed black lines). $T_c$ is determined from the peak of the susceptibility and specific heat.}
    \label{fig:universality}
\end{figure}
%
%

\begin{figure}[htp]
    \includegraphics[width=\columnwidth]{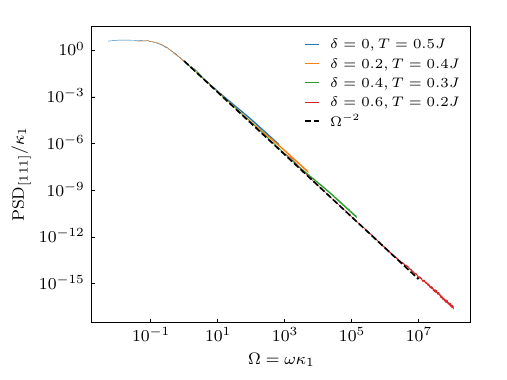}
    \caption{The magnetic noise power (PSD) of $\SL$ spins, for four values of $\delta$ in the regime $0 \le \delta < 1$. This plot is a more detailed version of the inset in panel (b) of \autoref{fig:PSD}. Plotting ${\rm PSD}_{[111]} / \kappa_1$ versus $\omega \kappa_1$ leads to the desired data collapse, as explained in the text. The black dashed line indicates an expected Lorentzian behavior for diffusion.}
    \label{fig:scaling_collapse}
\end{figure}
\section{Further numerical results \label{app:further_numerical}}
\subsection{Phase transition at $\delta > 1$ \label{app:ordering}}
We have identified a symmetry-breaking phase transition upon cooling when $\delta > 1$ (and $\delta < -3 $), \cf see \autoref{sec:GS and excitations} and \ref{sec:thermo}. This is a conventional paramagnet to ferromagnet phase transition in an Ising spin system, and we expect it to fall within the 3d Ising universality class~\cite{vojta2003quantum,jaubert2010spin}.
In the thermodynamic limit, one would then expect the susceptibility to diverge as $\chi \sim \left|T - T_c \right|^{-\gamma}$ and the specific heat to diverge as $C_v \sim \left|T- T_c \right|^{-\alpha}$ as $T$ approaches $T_c$ from above. 

In \autoref{fig:universality}, we plot the specific heat and the susceptibility for all spin subspecies for temperatures close to the critical temperature, approaching $T_c$ from above. The critical temperature is taken to be the temperature at which we observe the maximal specific heat and susceptibility value. We display results for two different linear system sizes, $L = 6$ and $10$, and find that the growth of both the susceptibility and specific heat approximately agree with the predicted critical exponents: $\gamma=1.23$ and $\alpha=0.11$ \cite{cardy1996scaling}. This is consistent with the thermal phase transitions being 3d Ising transitions, as expected.

%
%
\begin{figure*}[t]
    \centering
    \includegraphics[width=1\textwidth]{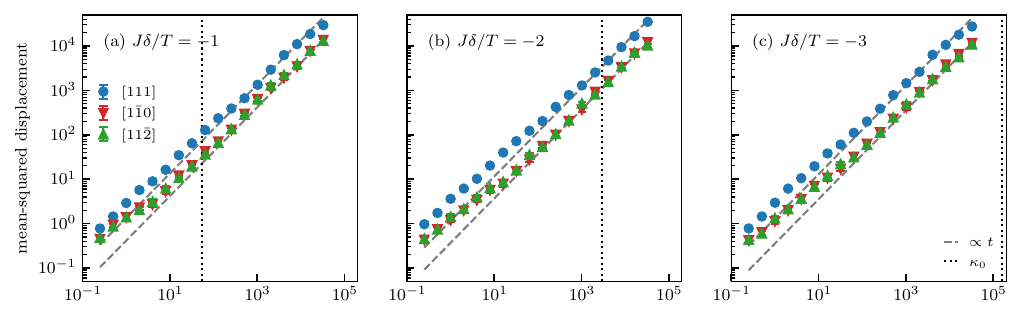}
    \includegraphics[width=1\textwidth]{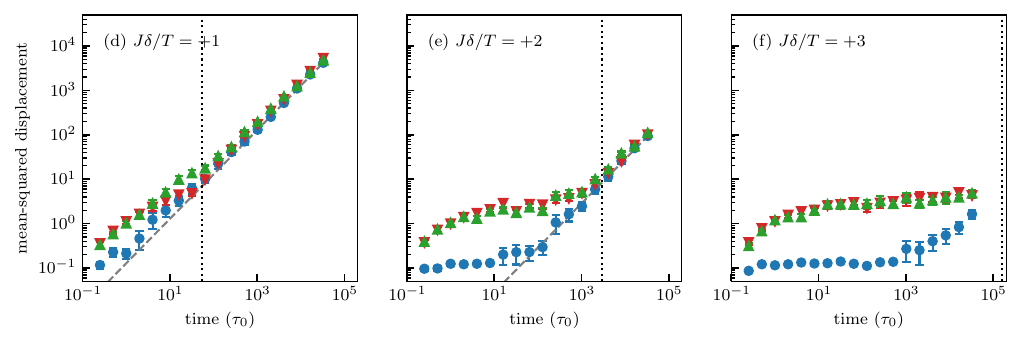}
    \caption{The MSD of individual isolated, meaning no annihilation or creation, monopoles measured along the $[111]$ direction (blue circles) and along two orthogonal directions: $[1\bar{1}0]$ (red down triangles) and $[11\bar{2}]$ (green up triangles). The MSD measured in units of the squared distance between the center of two neighboring tetrahedra is shown for six different values of $J \delta/T$, with negative $\delta$ in the upper row and positive $\delta$ in the lower row. The dashed gray lines indicate linear growth in time associated with diffusive monopole motion. The vertical dotted black lines indicate $\kappa_0 = {\rm e}^{4 J \delta / T}\tau_0 $. In the $\delta > 0$ regime, one expects monopoles to be energetically localized to small clusters on timescales smaller than this.}
    \label{fig:MSDsupp}
\end{figure*}
%
%
\subsection{Scaling collapse for type 1A monopoles}\label{ssec:scaling}
In \autoref{sec:dynamics}, we discussed the dynamical behavior of type 1A monopoles and argued that the magnetic noise of the $\SL$ spins is completely dominated by their motion. Remember that type 1B monopoles cannot move through spins on $\SL$. We also argued that type 1A monopoles move diffusively in the entire spin ice regime, and we, therefore, expect them to generate a Lorentzian magnetic noise spectrum \cite{hallen2022}, described by
\begin{equation}
    f=\frac{A \tau}{1+(\omega \tau)^2} 
    \, ,
    \label{eq:Lorentzian_1}
\end{equation}
where $\tau$ is the correlation time. The fact that the diffusion constant is anisotropic in the $-1 < \delta < 0$ regime does not change the frequency dependence of the noise but impacts the magnitude of noise measured along different crystallographic axes.

The correlation time of the noise is expected to be inversely proportional to the density of the dynamically active monopoles, in this case, the type 1A monopoles. 
We have implicitly made a scaling collapse in panel (a) of \autoref{fig:PSD} for $-1 < \delta \le 0$, since we chose values of $T$ and $\delta$ which keep $(2-2\vert \delta \vert)J/T$ constant. The density of type 1A monopoles goes as $\exp\left((2-2\vert \delta \vert)J/T\right)$ for negative $\delta$.

In the regime $0 < \delta < 1$ we thus expect the PSD of spins on $\SL$ to go as 
\begin{equation}
    {\rm PSD}_{[111]} \approx A \frac{\kappa_1}{1 + \left(\omega \kappa_1  \right)^2}
    \, ,
\end{equation}
with a global constant $A$ that is the same for all $T$ and $\delta$ ($A=2$ for our parametrization) and
\begin{equation}
    \kappa_1 = e^{-\left(2 + 2\vert\delta\vert \right) J / T} \tau_0
    \, 
\end{equation}
from \autoref{eq:kappa1}.
Indeed, we find a perfect scaling collapse for ${\rm PSD}_{[111]}$ curves at different $T$ and $\delta$ by plotting ${\rm PSD}_{[111]} / \kappa_1$ versus $\kappa_1 \omega$. This is shown in \autoref{fig:scaling_collapse} for the regime $0 \le \delta < 1$.

%
%
\begin{figure*}
    \centering
    \includegraphics[width=\textwidth]{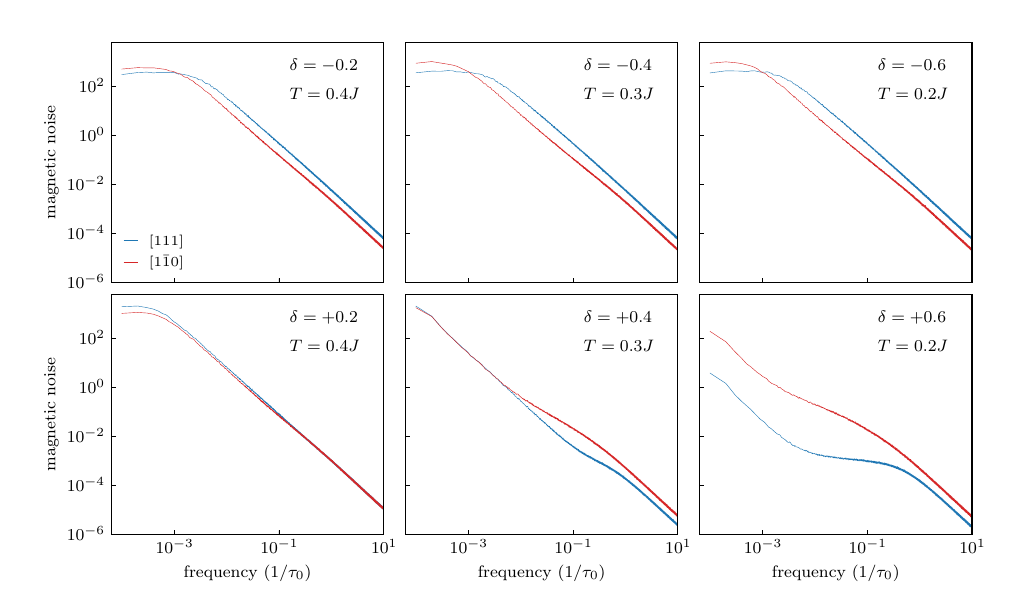}
    \caption{Additional magnetic noise (PSD) curves, measured along the $[111]$ direction (blue) and $[1\bar{1}0]$ (red), with negative $\delta$ in the upper row and positive $\delta$ in the lower row. The $[1\bar{1}0]$ direction is arbitrarily chosen from a set of directions orthogonal to $[111]$; the choice is confirmed to have no effect on the observed PSD. Unlike the PSD measured for individual spin subspecies, as shown in \autoref{fig:PSD}, the directional PSD is observable experimentally. At $\delta < 0 $, the $[1\bar{1}0]$ noise is `quieter' than $[111]$ and the opposite is true for $\delta>0$ due to different dynamical constraints imposed on type 1A and 1B monopoles. 
    The lifting of localization and onset of diffusive behavior at long timescales for monopoles when $\delta > 0$ appears as a $\omega^{-2}$ decay of the PSD at low frequencies.
    Note, however, that the exact agreement at low frequency of the two PSD curves for $\delta=0.4,\ T=0.3J$ is coincidental and only occurs at (in this case accidentally) fine-tuned choices of parameters.
    }
    \label{fig:PSD_direction}
\end{figure*}
%
%
\subsection{Further MSD and PSD results}
Here, we present further numerical results on the MSD of monopoles and directional magnetic noise under various settings of $\delta$ and $T$, complementary to the main text. \autoref{fig:MSDsupp} shows how the behavior of MSD depends on different values of $J\delta/T$. The monopole movement in the positive $\delta$ regime is much more constrained, as explained in the main text. 

The PSDs shown in \autoref{fig:PSD_direction} are measured along two orthogonal directions, providing a different perspective from the sublattice resolved PSD curves presented in the main text. The PSD along $[111]$ has a mixed contribution from monopole movement through $\SL$ and $\SLa$ as spins belonging to $\SLa$ have nonzero components along the $[111]$ direction, while the PSD along $[1\bar{1}0]$ has contributions from $\SLa$ spins only. The anisotropic diffusion of type 1A monopoles for $\delta < 0$ is manifested due to dynamical constraints imposed in directions other than $[111]$, whereas signals of dynamical localization appear in the suppression of $[111]$ directional PSDs for type 1B monopoles at $\delta > 0$.

The PSD measured along individual lattice directions is the experimentally relevant quantity, while it also demonstrates a clearer connection with the monopole MSD results we present. We notice the similarities of the anisotropic-to-isotropic transition in time observed in both \autoref{fig:MSDsupp} and \autoref{fig:PSD_direction} for $\delta > 0$.

\end{document}